\documentclass[11pt,a4paper]{article}
\pdfoutput=1

\usepackage{jheppub}

\usepackage{amssymb, amsthm}
\usepackage{amsmath}    
\usepackage{mathrsfs}   
\usepackage{graphicx}   
\usepackage{color}      
\usepackage{footnote}   

\usepackage{slashed}    
\usepackage{subfigure}  
\usepackage{hyperref}   

\usepackage{rotating}   
\usepackage{multirow}   
\usepackage[normalem]{ulem} 
\usepackage[utf8]{inputenc}

\newcommand{\beq}{\begin{equation}}
\newcommand{\eeq}{\end{equation}}

\newcommand{\nn}{\nonumber}

\def \t {\tilde}

\subheader{
\begin{flushright}
{\small
{TUM-HEP-1417/22, JLAB-THY-22-3720}
}
\end{flushright}
}

\begin{document}

\title{Gluon Transverse-Momentum-Dependent Distributions from Large-Momentum Effective Theory}

\author[a]{Ruilin Zhu,}
\emailAdd{rlzhu@njnu.edu.cn}
\affiliation[a]{Department of Physics and Institute of Theoretical Physics,
Nanjing Normal University, Nanjing, Jiangsu 210023, China}

\author[b,1]{Yao Ji \note{Corresponding author.},}
\emailAdd{yao.ji@tum.de}
\affiliation[b]{Physik Department T31, James-Franck-Stra\ss e 1, 
   Technische Universit{\"a}t M{\"u}nchen,\\
   D-85748 Garching, Germany}

\author[c,d,2]{Jian-Hui Zhang \note{Corresponding author.},}
\emailAdd{zhangjianhui@cuhk.edu.cn}
\affiliation[c]{School of Science and Engineering, The Chinese University of Hong Kong, Shenzhen 518172, China}
\affiliation[d]{Center of Advanced Quantum Studies, Department of Physics, Beijing Normal University, Beijing 100875, China}

\author[e,f,3]{and Shuai Zhao \note{Corresponding author.}}
\emailAdd{szhao@odu.edu}
\affiliation[e]{Department of Physics, Old Dominion University,\\
	 4600 Elkhorn Ave., Norfolk, VA 23529, USA}
\affiliation[f]{Theory Center, Thomas Jefferson National Accelerator Facility,\\ 12000 Jefferson Ave., Newport News, VA 23606, USA}


\abstract{We demonstrate that gluon transverse-momentum-dependent parton distribution functions (TMDPDFs) can be extracted from lattice calculations of appropriate Euclidean correlations in large-momentum effective theory (LaMET). Based on perturbative calculations of gluon unpolarized and helicity TMDPDFs, we present a matching formula connecting them and their LaMET counterparts, where the latter are renormalized in a scheme facilitating lattice calculations and converted to the $\overline{\rm MS}$ scheme. The hard matching kernel is given up to one-loop level. We also show that the perturbative result is independent of the prescription used for the pinch-pole singularity in the relevant correlations. Our results offer a guidance for the extraction of gluon TMDPDFs from lattice simulations, and have the potential to greatly facilitate perturbative calculations of the hard matching kernel. }


\maketitle


\section{Introduction}
\label{sec:Introduction}
Transverse-momentum-dependent parton distribution functions (TMDPDFs) generalize the collinear PDFs to incorporate the transverse momentum dependence of partons inside the hadrons, and thus play a crucial role in characterizing the three-dimensional structure of hadrons. They are also important inputs for describing multi-scale, noninclusive observables at high-energy colliders such as the LHC~\cite{Constantinou:2020hdm}. Understanding the TMDPDFs has been an important goal of many experimental facilities around the world, such as COMPASS at CERN, JLab 12 GeV upgrade, RHIC, and in particular, the forthcoming Electron-Ion Collider in US and China. So far, our knowledge of TMDPDFs mainly comes from the measurements of semi-inclusive deep-inelastic scattering (SIDIS) and Drell-Yan processes, for which the TMD factorization has been proven to hold~\cite{Collins:2011zzd,Echevarria:2011epo,Echevarria:2012js}. Based on these data, various fittings have been carried out to extract the TMDPDFs (see, e.g., ~\cite{Bacchetta:2017gcc,Scimemi:2017etj,Bertone:2019nxa,Scimemi:2019cmh,Bacchetta:2019sam,Bacchetta:2020gko}). However, the SIDIS and Drell-Yan processes are primarily induced by quarks/antiquarks, the information that can be extracted from them is mostly about the quark TMDPDFs. Our knowledge of gluon TMDPDFs is still very limited. Despite that several processes have been proposed to extract gluon TMDPDFs (see,  e.g., Refs.~\cite{Zhu:2013yxa,Scarpa:2019fol,Bacchetta:2022crh} and references therein), we are lacking of experimental data on such processes. Therefore, theoretical calculations of the TMDPDFs, in particular of the gluon TMDPDFs, can play an important complementary role to phenomenological approaches.

Calculating the TMDPDFs from theory has been a long-standing challenge in hadron physics, primarily because they are nonperturbative quantities defined in terms of light-cone correlations. A few pioneer calculations were carried out based on the Lorentz invariance approach, see, e.g., \cite{Hagler:2009mb,Musch:2011er,Engelhardt:2015xja}. In the past few years, significant progress~\cite{Ji:2014hxa,Ji:2018hvs,Ji:2019sxk,Ji:2019ewn,Ji:2020jeb,Ebert:2018gzl,Ebert:2019okf,Ebert:2019tvc,Shanahan:2019zcq,Ebert:2020gxr,Vladimirov:2020ofp,Ji:2021uvr,Zhang:2022xuw,Ebert:2022fmh} has been made within the framework of large-momentum effective theory (LaMET)~\cite{Ji:2013dva,Ji:2014gla,Ji:2020ect}, which allows us to calculate the quark TMDPDFs from first-principles lattice QCD. The calculation begins with the so-called quark quasi-TMDPDFs defined as the hadron matrix elements of suitable quark bilinear operators with a staple-shaped gauge link of finite length along the spacelike direction. The finite link length regulates the pinch-pole singularity associated with infinitely long gauge links~\cite{Ji:2020ect}. By combining with the square root of a Euclidean Wilson loop and suitable short-distance matrix elements~\cite{Ji:2021uvr,Zhang:2022xuw}, the result is free of ultraviolet (UV) divergences. The renormalized quark quasi-TMDPDF can then be factorized into the standard TMDPDF associated with a perturbative hard kernel, a Collins-Soper evolution part and a ``reduced soft function", up to power suppressed contributions~\cite{Ji:2019sxk,Ebert:2019okf,Ji:2020ect}. So far, considerable attention~\cite{LatticeParton:2020uhz,Li:2021wvl,LPC:2022ibr,Shanahan:2020zxr,Shanahan:2021tst,Schlemmer:2021aij} has been paid to the lattice calculation of quark TMDPDFs, although the complete lattice result is not yet available. In contrast, much less is known about the calculation of gluon TMDPDFs. 

In this paper, we consider the extraction strategy of gluon TMDPDFs from lattice simulations~\footnote{While this work is in progress, another paper~\cite{Schindler:2022eva} appears, where a conclusion similar to ours has been obtained, but the formulas are slightly different due to different choices of gluon quasi-TMDPDF operators.}. Among all eight leading-twist gluon TMDPDFs, we take the unpolarized and helicity ones as an example, as their perturbative calculation does not require the transverse momentum of external states. Based on one-loop results of the gluon quasi-TMDPDFs and TMDPDFs, we demonstrate that the matching relation between them takes a similar form as in the quark case, and give the explicit expression of the perturbative matching kernel up to one-loop, where the gluon quasi-TMDPDFs are renormalized in a scheme that facilitates lattice calculations~\cite{Ji:2021uvr,Zhang:2022xuw}. Moreover, note that the quasi-TMDPDFs can be viewed, in a sense, as the definition of TMDPDFs with the hadron momentum being a rapidity regulator~\cite{Ji:2020ect}. The finite link length is then kept as a regulator of the pinch-pole singularity, and used in perturbative calculations in the literature. This makes such calculations very complicated. Actually, as will be seen below, the same result can be obtained by employing a different prescription for the pinch-pole singularity, e.g., the $\delta$-prescription
, in perturbative calculations, indicating independence of the result on the pinch-pole singularity regulator. This observation has the potential to greatly facilitate calculations of the hard matching kernel. 

The rest of the paper is organized as follows. In Sec.~\ref{sec:notations} we set up our notations and conventions. In Sec.~\ref{sec:gtmdpdfs} and Sec.~\ref{sec:gqtmdpdfs} we present the one-loop calculation for the gluon TMDPDFs and quasi-TMDPDFs, respectively. In Sec.~\ref{sec:schemeindep} we compare the results of two different prescriptions for the pinch-pole singularity, and discuss their implications. Sec.~\ref{sec:factor} contains the matching formula connecting the renormalized gluon quasi-TMDPDFs and TMDPDFs. We conclude in Sec.~\ref{sec:concl}. Some useful formulas are listed in the Appendix.

\section{Notations and conventions}
\label{sec:notations}
In this section, we set up our notations and conventions. To describe a fast-moving hadron, we introduce two light-cone vectors 
\begin{align}
n^\mu=\frac{1}{\sqrt{2}}(1,\vec 0_\perp,-1),~~~~\bar{n}^\mu=\frac{1}{\sqrt{2}}(1,\vec 0_\perp,1),
\end{align}
which satisfy $n^2=\bar{n}^2=0$ 
and $n\cdot \bar{n}=1$. Any four-vector can then be decomposed as
\begin{align}
k^\mu=(k^0,\vec k_\perp,k^z)=k^+ \bar{n}^\mu+k^-n^\mu+k_\perp^\mu,
\end{align}
where the light-cone components are expressed as $k^+=n\cdot k=\frac{1}{\sqrt 2}(k^0+ k^z)$ and $k^-=\bar{n}\cdot k=\frac{1}{\sqrt 2}(k^0- k^z)$.
Thus, the momentum of the fast-moving hadron $P^\mu=(P^0,\vec 0_\perp,P^z)$ can be rewritten as $P^\mu=P^+ \bar{n}^\mu+P^-n^\mu$ with $P^+\gg P^-=M^2/(2P^+)$, where $M$ is the hadron mass and is much smaller than the hadron momentum. 

To project out the transverse components, it is  useful to define the transverse metric and the anti-symmetric tensor as
\begin{align}
g_\perp^{\mu\nu}=g^{\mu\nu}-n^\mu\bar{n}^\nu-\bar{n}^\mu n^\nu,~~\varepsilon_\perp^{\mu\nu}=
n_\alpha\bar{n}_\beta \varepsilon^{\alpha\beta\mu\nu}.
\end{align}

\begin{figure}[th]
\begin{center}
\includegraphics[width=0.3\textwidth]{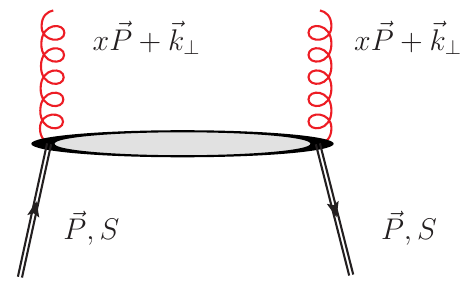}
\caption{ Schematic diagram of gluon TMDPDFs.}\label{fig:tmdplot}
\end{center}
\end{figure}

The gluon TMDPDFs are defined in terms of nonlocal gluon bilinear operators. A schematic diagram of them is shown in Fig.~\ref{fig:tmdplot}. To preserve gauge invariance,  a gauge link is inserted between the gluon fields at different spacetime points. For gluon TMDPDFs, the following gauge link $\cal W_{\pm\infty}$ is employed 
\begin{align}
{\cal W}_{\pm\infty}(b,a)={\cal W}^{\dagger}_n\left(b^{\mu},\pm\infty\right) {\cal W}^{\dagger}_T\left(\pm\infty n^\mu;b^{\mu}_{\perp},a^{\mu}_{\perp}\right) {\cal W}_n\left(a^\mu,\pm\infty\right),
\end{align}
where the gauge links in any light-cone $n_i$-direction and transverse direction are given respectively by
\begin{align}
\mathcal{W}_{n_i}\left(a^{\mu},\pm\infty\right)=&{\cal P}e^{\mp ig\int_0^\infty d s n_i\cdot A(a^{\mu}+sn_i^\mu)},\\
\mathcal{W}_T\left(y^{\mu};b^{\mu}_{\perp},a^{\mu}_{\perp}\right)=&{\cal P}e^{-ig\int_{\vec{a}_\perp}^{\vec{b}_\perp} d \vec{s}_T \cdot\vec{ A}_T(y^{\mu}+s^\mu_T)},
\end{align}
and they are defined in the adjoint representation.

Analogously, for gluon quasi-TMDPDFs that can be accessed from Euclidean lattice QCD, one chooses the following spatial gauge link $\widetilde{\cal W}_{\pm L}$ with a finite longitudinal length $L$
\begin{align}
\widetilde{\cal W}_{\pm L}(b,a)=\widetilde{\cal W}^{\dagger}_{v}\left(b^{\mu},\pm L\right) \widetilde{\cal W}^{\dagger}_T\left(\pm L v^\mu;b^{\mu}_{\perp},a^{\mu}_{\perp}\right) \widetilde{\cal W}_v\left(a^\mu,\pm L\right),\label{gl-quasi}
\end{align}
where the gauge links in any off light-cone $v_i$-direction and transverse direction
are 
\begin{align}
\widetilde{\mathcal W}_{v_i}\left(a^{\mu},\pm L\right)=&{\cal P}e^{\mp ig\int_0^L d s v_i\cdot A(a^{\mu}+sv_i^\mu)},\\
\widetilde{\mathcal W}_T\left(y^{\mu};b^{\mu}_{\perp},a^{\mu}_{\perp}\right)=&{\mathcal W}_T\left(y^{\mu};b^{\mu}_{\perp},a^{\mu}_{\perp}\right).
\end{align}
For later convenience, we have already introduced two new off-light-cone vectors
\begin{align}
v^\mu=(0,\vec 0_\perp,-1),~~~~\bar{v}^\mu=(1,\vec 0_\perp,0). \label{eq:def:v}
\end{align}

\section{Gluon TMDPDFs at one-loop}
\label{sec:gtmdpdfs}

\subsection{Definition}
To study the leading-twist gluon TMDPDFs, we begin with the following non-local correlator 
\beq\label{eq:gcorrfunc}
\Phi^{\mu\nu,\rho\sigma}(\xi,P,S)=\left\langle PS\left|F^{\mu\nu}\left(\frac \xi 2\right){\cal W}_{-\infty}\left(\frac \xi 2,-\frac \xi 2\right)F^{\rho\sigma}\left (-\frac {\xi} 2\right)\right|PS\right\rangle,
\eeq
where we have taken the past-pointing gauge link ${\cal W}_{-\infty}$ as an example, the analysis of the future-pointing case is similar. $|PS\rangle$ denotes an external hadron with momentum $P^\mu=P^+\bar{n}^\mu+P^- n^\mu$ and 
 spin $S^\mu=S_L(P^+\bar{n}^\mu-P^- n^\mu)/(2M)+S_\perp^\mu$ with $S\cdot P=0$ and $S^2=-1$. The two gluon field tensors are separated with $\xi^\mu=(0,b^-,\vec b_\perp)$, and $b^\pm=1/\sqrt 2(b^0\pm b^z)$ are the light-cone coordinates.

The Fourier transform of Eq.~(\ref{eq:gcorrfunc}) 
\beq
\Phi^{\mu\nu,\rho\sigma}(x,\vec k_\perp,S)=\frac{1}{x n\cdot P}\int\frac{db^- d^2\vec b_\perp}{(2\pi)^3}e^{-ik\cdot \xi}\Phi^{\mu\nu,\rho\sigma}(\xi,P,S)
\eeq
defines eight leading-twist gluon TMDPDFs~\cite{Mulders:2000sh,Meissner:2007rx}
\begin{align}
\Phi^{n i, n i}(x,\vec k_\perp,S)&=f_1^g(x, \vec k_\perp^2)-\frac{\epsilon^{ij}_\perp k_\perp^i S_\perp^j}{M}f_{1T}^{\perp g}(x,\vec k_\perp^2),\nn\\
i\epsilon_\perp^{ij}\Phi^{n i, n j}(x,\vec k_\perp,S)&=\lambda g_{1L}^g(x, \vec k_\perp^2)+\frac{k_\perp^i S_\perp^i}{M}g_{1T}^{g}(x,\vec k_\perp^2),\nn\\
-\hat S\Phi^{n i, n j}(x,\vec k_\perp,S)&=-\frac{\hat S k_\perp^i k_\perp^j}{2M^2}h_1^{\perp g}(x, \vec k_\perp^2)+\frac{\lambda \hat S \epsilon_\perp^{jk}k_\perp^i k_\perp^k}{2M^2}h_{1L}^{\perp g}(x, \vec k_\perp^2)+\frac{ \hat S \epsilon_\perp^{jk}k_\perp^i S_\perp^k}{2M}h_{1T}^{g}(x, \vec k_\perp^2)\nn\\
&+\frac{ \hat S \epsilon_\perp^{jk}k_\perp^i k_\perp^k k_\perp^m S_\perp^m}{2M^3}  h_{1T}^{\perp g}(x, \vec k_\perp^2),
\end{align}
where $k^+=x P^+$, and $\hat S$ denotes the symmetric and traceless part of the correlator. Throughout this paper, we consider the unpolarized and helicity TMDPDFs $f_1^g$ and $g^g_{1L}$, as their perturbative calculation does not require transverse momentum of the external state. The remaining TMDPDFs will be considered in future work.

The TMDPDFs defined above contains rapidity divergences associated with infinitely-long light-like gauge links, which need to be regularized by introducing an appropriate regulator. Various rapidity regulators have been proposed in the literature (for a summary see Ref.~\cite{Ebert:2019okf}). Here we follow Ref.~\cite{Echevarria:2011epo} to adopt the $\delta$-regulator. The rapidity divergences in the TMDPDFs then appear as logarithms of $\delta$. Such divergences can be removed by introducing a soft function which is defined as the vacuum expectation value of the following Wilson loop
\begin{eqnarray}
S(\vec{b}_{\perp}, \mu)
&=&\frac{1}{N_{c}^{2}-1}\left\langle 0\left|{\mathcal W}^{\dagger}_{\bar{n}}\left(\frac{b_{\perp}^{\mu}}{2},-\infty\right) {\mathcal W}^{\dagger}_T\left(-\infty {\bar{n}}^\mu;\frac{b_{\perp}^{\mu}}{2},-\frac{b_{\perp}^{\mu}}{2}\right) {\mathcal W}^{\dagger}_{\bar{n}}\left(-\frac{b_{\perp}^{\mu}}{2},-\infty\right)\right.\right. \nonumber\\&&\left.\left.\times{\mathcal W}_n\left(\frac{b_{\perp}^{\mu}}{2},-\infty\right){\mathcal W}_T\left(-\infty n^\mu;\frac{b_{\perp}^{\mu}}{2},-\frac{b_{\perp}^{\mu}}{2}\right) {\mathcal W}^{\dagger}_n\left(-\frac{b_{\perp}^{\mu}}{2},-\infty\right) \right|0\right\rangle, \label{eq:def:softfac}
\end{eqnarray}
where ${\mathcal W}_n$ and ${\mathcal W}_{\bar n}$ are the gauge links
including the soft gluon radiations along the light-cone directions $n$ and $\bar n$, respectively. $\mu$ is the renormalization scale in the $\overline{\rm MS}$ scheme. In the calculation below, we will work in the Feynman gauge. Thus, the transverse gauge links at infinity do not contribute~\cite{Belitsky:2002sm}. The physical TMDPDFs are then defined as
\beq
f_{\rm sub}(x,{\vec k}_\perp^2 ,\mu)=\int \frac{d^2\vec b_\perp}{(2\pi)^2}e^{i \vec k_\perp\cdot \vec b_\perp}f_{\rm sub}(x, \vec b_\perp^2 ,\mu).
\eeq
with
\begin{align}
f_{\rm sub}(x, \vec b_\perp^2 ,\mu)=\frac{f(x, \vec{b}_{\perp}^2, \mu)}{\sqrt{S(\vec{b}_{\perp}, \mu)}}.
\label{eq:fsub}
\end{align}
In the following, we will refer to $f$ and $f_{\rm sub}$ as the unsubtracted and subtracted TMDPDFs, respectively.

\subsection{Unsubtracted TMDPDF}
Now we consider the perturbative calculation of the unsubtracted TMDPDF. To start with, we replace the hadron state $| P\rangle$ with an on-shell gluon state $| g (p)\rangle$ with $p$ denoting its momentum. Both the ultraviolet (UV) and infrared (IR) divergences are regularized with dimensional regularization with $D=4-2\epsilon$. 

\begin{figure}[thbp]
\begin{center}
\includegraphics[width=0.88\textwidth]{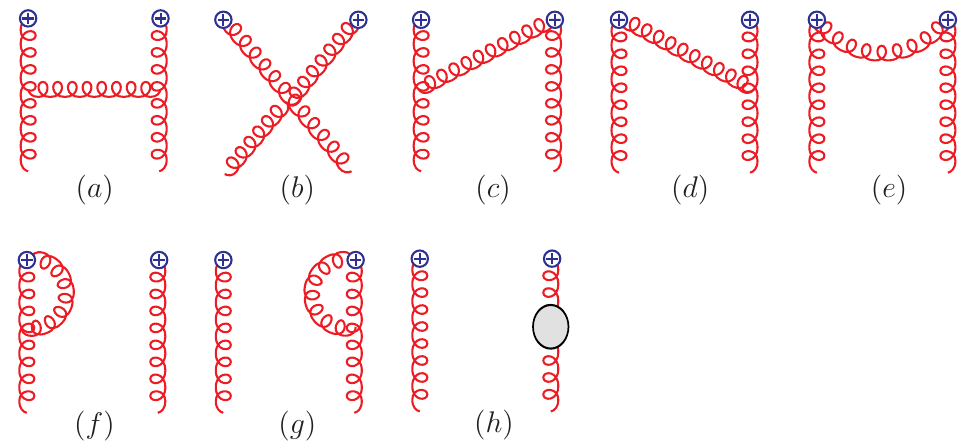}
\caption{ One-loop Feynman diagrams for the gluon TMDPDFs without gauge link interactions.}  
\label{fig:gluonTMDNOGL}
\end{center}
\end{figure}

\begin{figure}[thbp]
\begin{center}
\includegraphics[width=0.88\textwidth]{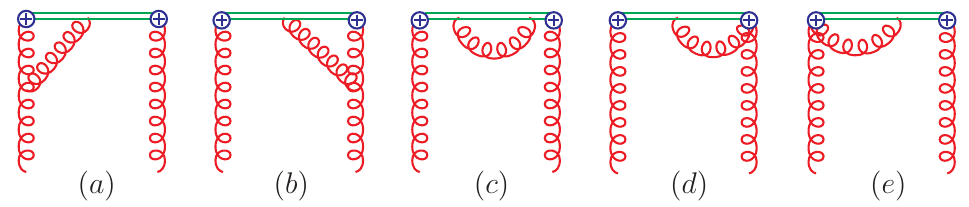}
\caption{ One-loop Feynman diagrams for the gluon TMDPDFs with gauge link interactions.}
\label{fig:gluonTMDGL}
\end{center}
\end{figure}

In Feynman gauge, the relevant one-loop diagrams for the unsubtracted gluon TMDPDFs are shown in Figs.~\ref{fig:gluonTMDNOGL} and \ref{fig:gluonTMDGL}. They can be categorized into two groups: the ones with and the ones without Wilson line interactions. In the following, we present the detailed calculation for the unpolarized gluon TMDPDF $f_1^g$. The calculation of $g_{1L}^g$ is similar, we will give the final result only.

Let us first look at the ladder diagram shown in Fig.~\ref{fig:gluonTMDNOGL}. 
The matrix element of the operator $F^{\alpha_1\beta_1} F^{\alpha_2 \beta_2}$ is given by
\begin{align}
&\left\langle p, \varepsilon^*\left|F^{\alpha_1\beta_1}\left(\frac{\xi}{2}\right) F^{\alpha_2 \beta_2}\left(-\frac{\xi}{2}\right)\right|p, \varepsilon \right\rangle\bigg\vert_{\ref{fig:gluonTMDNOGL}(a)}\nonumber\\
=&\tilde{\mu}^{2\epsilon}\int\frac{d^Dk}{(2\pi)^D} e^{i k\cdot \xi}(-gf_{a_1 b_1 c_1})[(p+k)^{\gamma_1} g^{\mu_1 \nu_1}+(-2k+p)^{\mu_1}g^{\nu_1 \gamma_1}+(k-2p)^{\nu_1}g^{\mu_1 \gamma_1}]\delta_{b_1 b_2}\nonumber\\
&\times(-gf_{a_2 c_1 b_c})
 [(p+k)^{\gamma_2} g^{\mu_2 \nu_2}+(p-2k)^{\mu_2}g^{\nu_2 \gamma_2}+(k-2p)^{\nu_2}g^{\mu_2 \gamma_2}]\nonumber\\
&\times(-i)(k^{\alpha_1}g^{\nu_1'\beta_1}-k^{\beta_1}g^{\nu_1'\alpha_1}) i(k^{\alpha_2} g^{\nu'_2\beta_2}
-k^{\beta_2} g^{\nu'_2\alpha_2})\nonumber\\
&\times\frac{-ig_{\nu_2 \nu'_2}}{k^2}\frac{-ig_{\nu_1
\nu'_1} }{k^2 }\frac{-ig_{\gamma_1
\gamma_2}}{(p-k)^2} \varepsilon_{\mu_1}\varepsilon^*_{\mu_2}\frac{\delta_{a_1 a_2}}{N_c^2-1},  \label{eq:gluonreal:amp1}
\end{align}
where $\tilde{\mu}\equiv\mu\sqrt{\frac{e^{\gamma_E}}{4\pi}}$ and $\gamma_E$ is the Euler constant.  
The vector $\xi$ can be decomposed as $\xi=b^- n+\vec{b}_{\perp}$. 
$\varepsilon$ and $\varepsilon^*$ are the polarization vectors of the initial and final states, respectively. The gluons are on-shell and transverse, i.e., $p^2=0$, $p\cdot \varepsilon=0$, so the matrix element above is gauge invariant. The unpolarized gluon TMDPDF is obtained by making the replacement $\varepsilon_{\mu_1}\varepsilon_{\mu_2}^*\to -g_{\perp,\mu_1\mu_2}/{(D-2)}$.
A direct calculation yields 
\begin{align}
  x f_1^{g,(1)} (x, \vec{k}_\perp^2 )|_{\ref{fig:gluonTMDNOGL}(a)}&=  \frac{\alpha_s C_A}{2\pi^2} \frac{(2\pi\tilde{\mu})^{2\epsilon}}{\vec{k}_\perp^2}(2-2x+3x^2-2x^3) \theta(0<x<1).
\end{align}
We note the support of the TMDPDF is $0<x<1$. In the following, we omit the Heaviside theta functions in our expressions for simplicity.

The calculations of the remaining diagrams are similar, and the results read
\begin{align}
x f_1^{g,(1)} (x,\vec {k}_{\perp}^2)|_{\ref{fig:gluonTMDNOGL}(c)}&=x f_1^{g,(1)} (x,\vec {k}_{\perp}^2 )|_{\ref{fig:gluonTMDNOGL}(d)}=	-\frac{\alpha_s C_A}{4\pi^2 } (2\pi\tilde{\mu})^{2\epsilon}\frac{1}{\vec{k}_{\perp}^2}x(1+x) ,\\
x f_1^{g,(1)} (x,\vec {k}_{\perp}^2 )|_{\ref{fig:gluonTMDNOGL}(f)}&=x f_1^{g,(1)} (x,\vec {k}_{\perp}^2)|_{\ref{fig:gluonTMDNOGL}(g)}=-\frac{3 \alpha_s C_A}{8\pi}\delta(1-x)\delta^2(\vec{k}_{\perp})\left(\frac{1}{\epsilon_{\mathrm{UV}}}-\frac{1}{\epsilon_{\mathrm{IR}}}+\ln\frac{\mu_{\mathrm{UV}}^2}{\mu_{\mathrm{IR}}^2} \right).
\end{align}
The contribution from gluon self energy diagrams is
\begin{align}
x f_1^{g,(1)} (x,
\vec{k}_{\perp}^2 ) \bigg\vert_{\ref{fig:gluonTMDNOGL}(h)+h.c.}=&\frac{\alpha_s
}{\pi}\delta(1-x) \delta^{(2)}(\vec{k}_{ \perp}) \left(\frac{5}{12}C_A-\frac13 T_F n_f\right) \left(\frac{1}{\epsilon_{\mathrm{UV}}}-\frac{1}{\epsilon_{\mathrm{IR}}}+\ln\frac{\mu_{\mathrm{UV}}^2}{\mu_{\mathrm{IR}}^2}\right) ,
\end{align}
where $C_A=N_c$ is the number of colors, $T_F=1/2$ and $n_f$ is the number of quark flavors.
There are no contributions from Fig.~\ref{fig:gluonTMDNOGL} (b) and Fig.~\ref{fig:gluonTMDNOGL} (e).

Fig.~\ref{fig:gluonTMDGL} contains the diagrams with gauge link interactions which are more interesting, since the Wilson lines along the lightcone direction lead to rapidity divergences. 
To regularize these divergences, we adopt the $\delta$-regularization~\cite{Echevarria:2015usa,Echevarria:2015byo}, in which the Wilson line propagator becomes
\begin{align}
    \frac{i}{k^{\pm}\pm i 0}\to \frac{i}{k^{\pm}\pm i \delta^{\pm}}\, .
\end{align}
Both the unsubtracted TMDPDF and the soft function depend on $\delta$, but the regulator dependence shall cancel out in the ratio in Eq.~(\ref{eq:fsub}), as we will see below. 

For Fig.~\ref{fig:gluonTMDGL} (a) and (b), we have  
\begin{align}
&x f_1^{g,(1)}(x,
\vec{k}_{\perp}^2 ) \bigg\vert_{\ref{fig:gluonTMDGL}(a)+(b)} \nonumber\\
=&	\frac{\alpha_s C_A}{4\pi^2} (2\pi\mu)^{2\epsilon}\frac{1}{\vec{k}_{\perp}^2} \bigg[ \bigg(\frac{x(1+x)}{1-x}\bigg)_+ +\delta(1-x)\left(-2\ln\frac{\delta^+}{p^+}+i\pi-\frac52\right)\bigg]\nonumber\\
&	-\frac{\alpha_s C_A}{4\pi}\delta(1-x)\delta^2(\vec{k}_{\perp}) \left(\frac{1}{\epsilon_{\mathrm{UV}}}-\frac{1}{\epsilon_{\mathrm{IR}}}+\ln\frac{\mu_{\mathrm{UV}}^2}{\mu_{\mathrm{IR}}^2}\right)\left(-2\ln\frac{\delta^+}{p^+}+i \pi -\frac52 \right)+h.c., 
\end{align}
whereas the contributions of Fig.~\ref{fig:gluonTMDGL} (c-e) are zero.

The above results are given in $(x, \vec k_\perp)$ space. They can also be Fourier transformed into $(x, \vec b_{\perp})$ space with the help of the formulas in the Appendix, and the total result for the unsubtracted unpolarized gluon TMDPDF in $(x, \vec b_{\perp})$ space is 
\begin{align}
x f_1^{g}(x,
\vec{b}_{\perp}^2)&=\delta(1-x)+\frac{\alpha_{s}}{2 \pi}C_A\left\{  -\left(\frac{1}{\epsilon_{\mathrm{IR}}}+\ln\frac{\mu^2 \vec{b}_{\perp}^2  e^{2\gamma_E}}{4}\right)x P_{gg} \right. \nonumber\\
&+\left.\left(\frac{1}{\epsilon_{\mathrm{UV}}}+\ln\frac{\mu^2 \vec{b}_{\perp}^2  e^{2\gamma_E}}{4}\right) \bigg(2 \ln \frac{\delta^+}{p^+}+\frac{\beta_0}{2 C_A}\bigg)\delta(1-x)\right\},
\end{align}
where we have set $\mu_{\mathrm{UV}}= \mu_{\mathrm{IR}}=\mu$, and 
$\beta_0=\frac{11C_A-4T_F n_f}{3}$. $ P_{g g}$ is the gluon-gluon splitting kernel
\begin{align}
 P_{g  g}=2C_A\left[\frac{x}{(1-x)_+}+\frac{1-x}{x}+x(1-x)\right]+\frac{\beta_0}{2}\delta(1-x).
\end{align}
From the above expression, one can see that the rapidity divergence exhibits itself as logarithms of $\delta^+$. To cancel this singularity, we need to include the contribution of the soft function.
\begin{figure}[th]
\begin{center}
\includegraphics[width=0.7\textwidth]{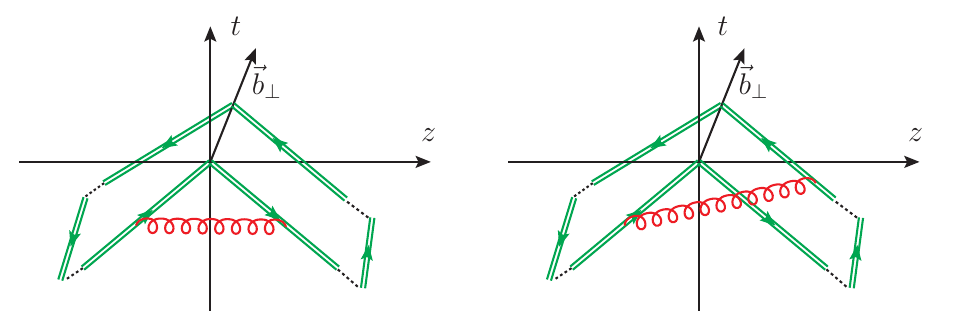}
\caption{ One-loop Feynman diagrams for the soft function, where the double-lines denote gauge links.}\label{fig:soft}
\end{center}
\end{figure}

\subsection{Soft function and subtracted TMDPDF}
In this subsection, we calculate the soft function defined in Eq.~\eqref{eq:def:softfac}. The relevant one-loop Feynman diagrams are plotted in Fig.~\ref{fig:soft}.
At one-loop, the calculation of the soft function required for the gluon TMDPDFs is essentially the same as that of the soft function required for the quark TMDPDFs, except that the color factors are different. Here we first calculate in $\vec k_\perp$ space and then Fourier transform to $\vec b_\perp$ space. The contribution of Fig.~\ref{fig:soft} (a) and its conjugate diagram is
\begin{align}
&S(
\vec{k}_{\perp}, \mu, \delta^+,\delta^-)\bigg\vert_{\mathrm{Fig}.\ref{fig:soft} (a+conj.)} \nonumber\\
=&-\frac{\alpha_{s} C_{A}}{2 \pi}\delta^{(2)}\left(\vec{k}_{ \perp}\right)\left(\frac{2}{\epsilon_{\mathrm{UV}}^{2}}-\frac{2}{\epsilon_{\mathrm{UV}}} \ln \frac{2\delta^{+} \delta^{-}}{\mu_{\mathrm{UV}}^{2}}+\ln ^{2} \frac{2\delta^{+} \delta^{-}}{\mu_{\mathrm{UV}}^{2}}+\frac{\pi^{2}}{2}\right).
\end{align}
Again, the $\delta$-regularization is used to deal with rapidity divergences. 
Here we choose $\delta^+$ and $\delta^-$ to regularize the rapidity divergence along the lightcone directions $n$ and $\bar{n}$, respectively. The double UV pole $1/\epsilon^2_{\mathrm{UV}}$ is a reflection of the cusp singularity.

Fig.~\ref{fig:soft} (b) and its conjugate diagram yield
\begin{eqnarray}
&&S(
\vec{k}_{\perp}, \mu, \delta^+, \delta^-)\bigg\vert_{\mathrm{Fig}.\ref{fig:soft} (b+conj.)} 
=-\frac{\alpha_{s} C_{A}}{\pi^{2}} \frac{(2\pi\tilde{\mu})^{2-2\epsilon} }{\vec{k}_{\perp}^{2}-2\delta^+\delta^-} \ln \frac{2\delta^{+} \delta^{-}}{\vec{k}_{\perp}^{2}}. 
\end{eqnarray}
Adding the two contributions together and doing the Fourier transform, we obtain the soft function in $\vec b_\perp$ space
\begin{align}
S(
\vec{b}_\perp, \mu, \delta^+, \delta^-)
=&1+\frac{\alpha_{s} C_{A}}{2 \pi} \bigg(-\frac{2}{\epsilon_{\mathrm{UV}}^{2}}+\frac{2}{\epsilon_{\mathrm{UV}}}\ln \frac{2\delta^{+} \delta^{-}}{\mu^{2}}+\ln^2\frac{\mu^2 \vec{b}_{\perp}^2  e^{2\gamma_E}}{4} \nonumber\\
&+2\ln\frac{\mu^2 \vec{b}_{\perp}^2  e^{2\gamma_E}}{4}\ln \frac{2\delta^{+} \delta^{-}}{\mu^{2}}+\frac{\pi^{2}}{6}\bigg).
\end{align}
It differs from the soft function in the fundamental representation~\cite{Echevarria:2012js} only by a color factor. As one can see from the expression above, the rapidity-regulator-dependent terms have the same structure as that in the unsubtracted gluon TMDPDF. By combining them, we have the final result for the subtracted gluon TMDPDF  
\begin{align}
 &x f_{1, \rm{sub}}^{g} \left( x, \vec{b}_{\perp}^2, \epsilon, \zeta\right)=  \delta(1-x)+\frac{\alpha_{s}}{2 \pi}\bigg\{ -\left(\frac{1}{\epsilon_{\mathrm{IR}}}+\ln\frac{\mu^2 \vec{b}_{\perp}^2  e^{2\gamma_E}}{4} \right) x  P_{g  g}\nonumber\\
&~~~~~~+C_{A} \bigg[\frac{1}{\epsilon_{\mathrm{UV}}^2}+ \left(\frac{1}{\epsilon_{\mathrm{UV}}}+\ln\frac{\mu^2 \vec{b}_{\perp}^2  e^{2\gamma_E}}{4} \right) \left( \frac{\beta_{0}}{2 C_A}+ \ln \frac{\mu^{2}}{\zeta} \right)\nonumber\\
&~~~~~~~~~~~~ -\frac{1}{2}\ln^2\frac{\mu^2 \vec{b}_{\perp}^2  e^{2\gamma_E}}{4}-\frac{\pi^{2}}{12}\bigg]\delta(1-x)\bigg\},
\end{align}
where $\zeta=2(p^+)^2\frac{\delta^-}{\delta^+}=2(p^+)^2=2(x P^+)^2$ with $\delta^-=\delta^+$ stands for the Collins-Soper scale. 
In the $\overline{\mathrm{MS}}$ scheme, the UV divergences can be removed by the renormalization factor
\begin{align}
    Z(\mu, \zeta, \epsilon)=1-\frac{\alpha_s}{2\pi} C_A\bigg[\frac{1}{\epsilon_{\mathrm{UV}}^2}+ \frac{1}{\epsilon_{\mathrm{UV}}} \left( \frac{\beta_{0}}{2 C_A}+ \ln \frac{\mu^{2}}{\zeta} \right)\bigg]\, .
\end{align}
Then the renormalized gluon TMDPDF reads
\begin{align}
   & x f_{1, \rm{sub}}^{g}\left(x, \vec{b}_{\perp}^2, \mu, \zeta\right)=  \delta(1-x)+\frac{\alpha_{s}(\mu)}{2 \pi}\bigg\{ -\left(\frac{1}{\epsilon_{\mathrm{IR}}}+\ln\frac{\mu^2 \vec{b}_{\perp}^2  e^{2\gamma_E}}{4} \right) x P_{g  g}\nonumber\\
&+C_{A} \left[  \ln\frac{\mu^2 \vec{b}_{\perp}^2  e^{2\gamma_E}}{4} \left( \frac{\beta_{0}}{2 C_A}+ \ln \frac{\mu^{2}}{\zeta} \right)-\frac{1}{2}\ln^2\frac{\mu^2 \vec{b}_{\perp}^2  e^{2\gamma_E}}{4}-\frac{\pi^{2}}{12}\right]\delta(1-x)\bigg\}\, . \label{eq:TMD:sub}
\end{align}

At last, we also calculate the polarized gluon TMDPDF $g_{1L}^g(x, \vec{k}_{\perp})$ at one-loop level, with external states being polarized gluon states. We make the replacement $\epsilon_{\mu_1}\epsilon^*_{\mu_2}\to -\frac{i}{2} \epsilon_{\perp,\mu_1\mu_2}$ in our calculation. The final result can be expressed as Eq.~\eqref{eq:TMD:sub} with the replacement $P_{g g}\to\Delta P_{g g}$, where 
\begin{align}
    \Delta P_{g g}= 2C_A\left[2(1-x)+\frac{x}{(1-x)_+}\right]+\frac{\beta_0}{2}\delta(1-x)
\end{align}
is the gluon-gluon splitting kernel for the polarized PDFs. 
\section{Gluon quasi-TMDPDFs at one-loop}
\label{sec:gqtmdpdfs}
\subsection{Definition}
The gluon quasi-TMDPDFs can be defined in terms of the following equal-time non-local correlator
\beq\label{eq:quasigcorrfunc}
\t\Phi^{\mu\nu,\rho\sigma}(\eta,P,S)=\lim_{L\to\infty}\frac{\langle PS|F^{\mu\nu}\left(\frac \eta 2\right)\widetilde{\cal W}_{-L}\left(\frac \eta 2,-\frac \eta 2\right)F^{\rho\sigma}\left (-\frac {\eta} 2\right)|PS\rangle}{\sqrt {Z_E(2L,|\vec b_\perp|)}},
\eeq
where the hadron momentum $P^\mu=(P^0,\vec 0_\perp,P^z)$ is finite, and  $\eta^\mu=(0,\vec b_\perp,b^z)$ is a purely spatial vector. $\widetilde{\cal W}_{-L}$ is the spatial gauge link of finite length inserted to ensure gauge invariance of the equal-time non-local correlator, which has been defined in Eq.~\ref{gl-quasi}.  

Note that in the above correlator, the hadron momentum is finite and the gauge link is along the spatial direction. Thus, no additional rapidity regulator is needed. The finite link length is introduced to regulate the so-called pinch-pole singularity associated with infinitely long Wilson lines. In a cutoff regularization such as lattice regularization, it is also associated with linear divergences from the Wilson lines. Nevertheless, the dependence on the length $L$ is cancelled by dividing by the square root of a Euclidean flat rectangular Wilson loop 
\begin{equation}
Z_E(2L,|\vec{b}_\perp|,\mu)=\frac{1}{N_{c}^{2}-1}\left\langle 0\left|\widetilde{\cal W}_{-2L}\left(\frac{b_\perp^\mu}{2}+L v^\mu,-\frac{b_\perp^\mu}{2}+L v^\mu \right)\mathcal{W}_T\left(L v^\mu;-\frac{b_\perp^\mu}{2},\frac{b_\perp^\mu}{2}\right) \right|0\right\rangle
\end{equation}
in Eq.~(\ref{eq:quasigcorrfunc}), so that we can take the limit $L\to\infty$. The introduction of $Z_E$ also removes additional contributions arising from the transverse gauge link.

In Ref.~\cite{Zhang:2018diq}, it has been identified that there are four gluon quasi-PDF operators that are multiplicatively renormalized~\cite{Zhang:2018diq,Wang:2019tgg}. Similar conclusion also applies to the gluon quasi-TMDPDF operators.  We find that the following operators can be renormalized multiplicatively
\begin{align}
    O^{(1)}&= F^{0i}\left(\frac{\eta}{2}\right)\widetilde{\cal W}\left(\frac{\eta}{2},-\frac{\eta}{2}\right)F^{0j}\left(-\frac{\eta}{2}\right),\\
    O^{(2)}&= F^{3i}\left(\frac{\eta}{2}\right)\widetilde{\cal W}\left(\frac{\eta}{2},-\frac{\eta}{2}\right)F^{3j}\left(-\frac{\eta}{2}\right),\\ 
    O^{(3)}&= \frac12 F^{0i}\left(\frac{\eta}{2}\right)\widetilde{\cal W}\left(\frac{\eta}{2},-\frac{\eta}{2}\right)F^{3j}\left(-\frac{\eta}{2}\right)+\frac12 F^{3i}\left(\frac{\eta}{2}\right)\widetilde{\cal W}\left(\frac{\eta}{2},-\frac{\eta}{2}\right)F^{0j}\left(-\frac{\eta}{2}\right),\\
    O^{(4)}&= F^{3\mu}\left(\frac{\eta}{2}\right)\widetilde{\cal W}\left(\frac{\eta}{2},-\frac{\eta}{2}\right)F^{3\nu}\left(-\frac{\eta}{2}\right), 
\end{align}
where $i, j$ denote the transverse components, while $\mu, \nu$ denote all four Lorentz components. For illustration purposes, we choose $O^{(1)}$ for our calculation~\footnote{In contrast, the authors of Ref.~\cite{Schindler:2022eva} choose the operator $F^{+i}(\frac{\eta}{2})\widetilde{\cal W}(\frac{\eta}{2},-\frac{\eta}{2})F^{+i}(-\frac{\eta}{2})$ which is, however, not multiplicatively renormalized.}. 
It also facilitates the nonperturbative renormalization to be discussed below. 

The Fourier transform of Eq.~(\ref{eq:quasigcorrfunc})
\beq
\t\Phi^{\mu\nu,\rho\sigma}(x,\vec k_\perp, S, P^z)=\frac{N}{x}\int\frac{db^z d^2\vec b_\perp}{(2\pi)^3}e^{-ik\cdot \eta}\t\Phi^{\mu\nu,\rho\sigma}(\eta,P,S)
\eeq
defines the quasi-TMDPDF counterparts of the gluon TMDPDFs, where $k^z=x P^z$, $N$ is a normalization factor depending on the choice of the operator. $N=v\cdot P/(\bar v\cdot P)^2=P^z/(P^0)^2$ for $O^{(1)}$ with $v$ and $\bar{v}$ being the spacelike and timelike vectors defined in Eq.~\eqref{eq:def:v}, and the relevant quasi-TMDPDFs are given by
\begin{align}
\t\Phi^{0 i, 0 i}(x,\vec k_\perp,S, P^z)&=\t f_1^g(x, \vec k_\perp^2, P^z)-\frac{\epsilon^{ij}_\perp k_\perp^i S_\perp^i}{M}\t f_{1T}^{\perp g}(x,\vec k_\perp^2, P^z),\\
i\epsilon_\perp^{ij}\t\Phi^{0 i, 0 j}(x,\vec k_\perp,S, P^z)&=\lambda \t g_{1L}^g(x, \vec k_\perp^2, P^z)+\frac{k_\perp^i S_\perp^i}{M}\t g_{1T}^{g}(x,\vec k_\perp^2, P^z),\\
-\hat S\t\Phi^{0 i, 0 j}(x,\vec k_\perp,S, P^z)&=-\frac{\hat S k_\perp^i k_\perp^j}{2M^2}\t h_1^{\perp g}(x, \vec k_\perp^2, P^z)+\frac{\lambda \hat S \epsilon_\perp^{jk}k_\perp^i k_\perp^k}{2M^2}\t h_{1L}^{\perp g}(x, \vec k_\perp^2, P^z)\nn\\
&+\frac{ \hat S \epsilon_\perp^{jk}k_\perp^i S_\perp^k}{2M}\t h_{1T}^{g}(x, \vec k_\perp^2, P^z) +\frac{ \hat S \epsilon_\perp^{jk}k_\perp^i k_\perp^k k_\perp^m S_\perp^m}{2M^3} \t h_{1T}^{g}(x, \vec k_\perp^2, P^z).
\end{align}


\subsection{Quasi-TMDPDF and renormalization}
\label{sec:gqtmdrenorm}
We begin with the unpolarized gluon quasi-TMDPDF defined by the operator $O^{(1)}$, which is given in $\eta$ space as
\begin{align}
\tilde h_1(\eta, P^z)=\frac{\langle P |F^{0i}(\frac{\eta}{2})\widetilde{\cal W}_{-L}(\frac{\eta}{2},-\frac{\eta}{2})F^{0i}(-\frac{\eta}{2})|P\rangle}{\sqrt{Z_E(2L,|\vec{b}_\perp|,\mu)}}.
\label{eq:quasi:def}
\end{align}
%


The calculation of the gluon quasi-TMDPDF is similar to that of the TMDPDF, and shares the same Feynman diagrams in Figs.~(\ref{fig:gluonTMDNOGL}-\ref{fig:gluonTMDGL}). For the diagrams without gauge links, we find the same result for the gluon quasi-TMDPDF as for the gluon TMDPDF at the leading power 
\begin{align}
x \t f_1^{g,(1)} (x,\vec {b}_{\perp}^2, p^z)|_{\ref{fig:gluonTMDNOGL}(a-h)}&=x f_1^{g,(1)} (x,\vec {b}_{\perp}^2)|_{\ref{fig:gluonTMDNOGL}(a-h)}. 
\end{align}
The quasi-TMDPDF outside the support $[0,1]$ is nonzero, but is power suppressed by $p^z$. Therefore, the quasi-TMDPDFs also have support $[0,1]$ at the leading power. For the sake of simplicity, we also omit the theta function $\theta(0<x<1)$ here.

The diagrams with gauge link interactions yield non-trivial contributions, which are
\begin{align}
&x \tilde{f}_1^{g, (1)}(x, \vec{k}_{\perp}^2, p^z)|_{\ref{fig:gluonTMDGL},(a)+(b)}\nonumber\\
=&	\frac{\alpha_s C_A}{2\pi }\bigg\{\frac{(2\pi\tilde{\mu})^{2\epsilon}}{\pi\vec{k}_{\perp}^2}\bigg[\frac{x(1+x)}{1-x} \bigg]_+ -\delta(1-x)\frac{(2\pi\tilde{\mu})^{2\epsilon}}{\pi\vec{k}_{\perp}^2}\bigg(\ln\frac{\vec{k}_{\perp}^2}{(2p^z)^2}+\frac52\bigg) \nonumber\\
	&+ 	    \delta(1-x) \delta^{2}(\vec{k}_{\perp}) \bigg[ \frac{1}{2}\left(\frac{1}{\epsilon_{\mathrm{UV}}}+\ln\frac{\mu_{\mathrm{UV}}^2}{(2p^z)^2}\right)-\frac{1}{\epsilon_{\mathrm{IR}}^2}-\frac{1}{\epsilon_{\mathrm{IR}}}\left(\ln\frac{\mu_{\mathrm{IR}}^2}{(2p^z)^2}+\frac52\right) \nonumber\\
	&-\frac12 \ln^2\frac{\mu_{\mathrm{IR}}^2}{(2p^z)^2}-\frac52\ln\frac{\mu_{\mathrm{IR}}^2}{(2p^z)^2}+\frac{\pi^2}{12}-4\bigg]\bigg\}\, .    
\end{align}
One can Fourier transform this result from ${\vec k}_{\perp}$-space to ${\vec b}_{\perp}$-space, with the help of the formulas listed in Appendix~\ref{Sec:FTint}. The result is 
\begin{align}
 &x \tilde{f}_1^{g,(1)}(x, \vec{b}_{\perp}^2, p^z)|_{\ref{fig:gluonTMDGL},(a)+(b)}  = 	\frac{\alpha_s C_A}{2\pi }  \bigg\{- \left(\frac{1}{\epsilon_{\mathrm{IR}}}+\ln\frac{\mu^2 \vec{b}_{\perp}^2 e^{2\gamma_E}}{4} \right)\bigg[\frac{x(1+x)}{1-x} \bigg]_+ \\
&+ \delta(1-x)     \left[ \frac{1}{2}\left(\frac{1}{\epsilon_{\mathrm{UV}}}+\ln\frac{\mu_{\mathrm{UV}}^2}{(2p^z)^2}\right) -\frac12\ln^2 ((p^z)^2\vec{b}_{\perp}^2 e^{2\gamma_E})+\frac52\ln ((p^z)^2\vec{b}_{\perp}^2 e^{2\gamma_E})-4\right] \bigg\}\nn.
\end{align}

\begin{figure}[th]
\begin{center}
\includegraphics[width=0.7\textwidth]{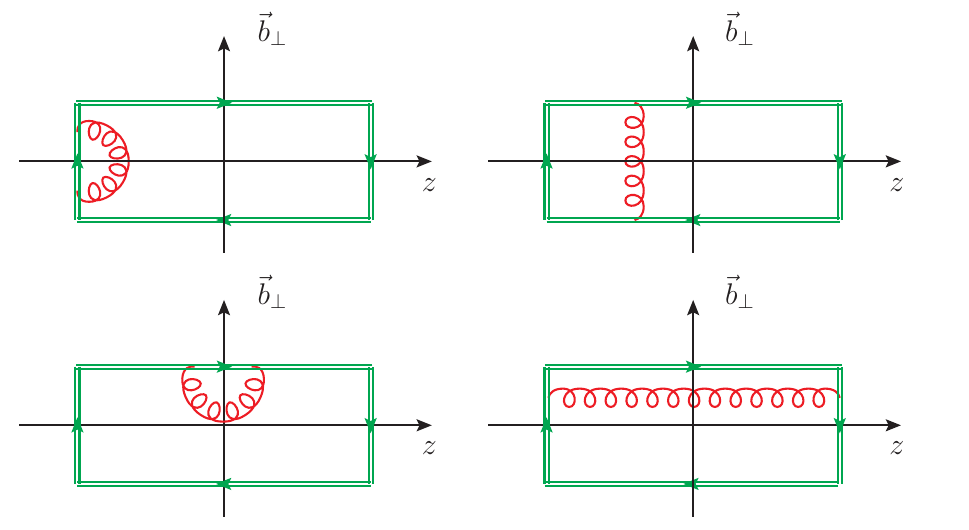}
\caption{One-loop Feynman diagrams for $Z_E(2L,\vec{b}_\perp,\mu)$ in the definition of quasi-TMDPDFs.}\label{fig:ZE}
\end{center}
\end{figure}

\noindent In ${\vec b}_{\perp}$ space, the collinear and soft poles in the $\delta(1-x)$ term, which appear as $1/\epsilon_{\mathrm{IR}}$ and $1/\epsilon_{\mathrm{IR}}^2$, are canceled.
This result can also be derived in coordinate space representation, by using the same trick as used in Ref.~\cite{Ebert:2019okf} for the quark quasi-TMDPDF calculation. 

There is no contribution from Fig.~\ref{fig:gluonTMDGL} (d) and (e). The self-energy of the Wilson line shown in Fig.~\ref{fig:gluonTMDGL}(c) is more conveniently calculated in ${\vec b}_{\perp}$ space, which gives
\begin{align}\label{eq:quasinum}
x \tilde{f}_1^{g,(1)}(x, \vec{b}_{\perp}^2, L)|_{\ref{fig:gluonTMDGL}(c)}=\frac{\alpha_s C_A}{2 \pi} \delta(1-x)\left(\frac{3}{\epsilon_{\mathrm{UV}}}+3\ln \frac{\mu_{\mathrm{UV}}^2 \vec{b}^2_\perp e^{2\gamma_E}}{4}+2+\frac{2\pi L}{|\vec{b}_\perp|}\right)\, .
\end{align}
It differs from the Wilson line self-energy in the fundamental representation only by a color factor. The pinch-pole singularity is regularized by $L$ as $L/|\vec{b}_{\perp}|$. 

The Feynman diagrams for the factor $Z_E(2L,\vec{b}_\perp,\mu)$ are shown in Fig.~\ref{fig:ZE}. The calculation is straightforward, and gives the following result
\begin{align}\label{eq:quasidenom}
 Z_E^{(1)}(2L,\vec{b}_\perp^2,\mu_{\mathrm{UV}})|_{\ref{fig:ZE}}=&\frac{2\alpha_s C_A }{\pi}\left(\frac{1}{\epsilon_{\mathrm{UV}}}+\ln \frac{\mu_{\mathrm{UV}}^2 \vec{b}^2_\perp e^{2\gamma_E}}{4}+1+\frac{\pi L}{|\vec{b}_\perp|}\right).
\end{align}
Again, it differs from the result in the fundamental representation only by a color factor. Note that the pinch-pole singularity cancels between Eq.~(\ref{eq:quasinum}) and the square root of $\sqrt {Z_E}$ at $O(\alpha_s)$.

In total, the gluon quasi-TMDPDF defined by operator $O^{(1)}$ reads
\begin{align}\label{Eq:O1result}
	x \tilde f_1^{g} (x,\vec {b}_{\perp}^2, p^z)=& x \tilde f_1^{g,(0)}(x,\vec {b}_{\perp}^2, p^z)+x \tilde f_1^{g,(1)}(x,\vec {b}_{\perp}^2, p^z)-\frac{1}{2}\delta(1-x)Z_E^{(1)}
	\nonumber\\
	=& \delta(1-x)+\frac{\alpha_s}{2 \pi}\bigg\{   -\bigg(\frac{1}{\epsilon_{\mathrm{IR}}}+\ln\frac{\mu^2 \vec{b}_{\perp}^2  e^{2\gamma_E}}{4} \bigg) x P_{g  g} \nonumber\\ 
	& +\delta(1-x) C_A \bigg[\bigg(
	\frac{\beta_0}{2C_A}-1\bigg)\bigg(\frac{1}{\epsilon_{\mathrm{UV}}}+\ln\frac{\mu^2 \vec{b}_{\perp}^2  e^{2\gamma_E}}{4}\bigg)-\frac{1}{2}\ln^2 ((p^z)^2 \vec{b}_{\perp}^2  e^{2\gamma_E}) \nonumber\\
	&+2\ln  ((p^z)^2 \vec{b}_{\perp}^2 e^{2\gamma_E})   -4\bigg] \bigg\}\, .
\end{align}

The above quasi-TMDPDFs still contain UV divergences that need to be renormalized. This is because, the square root of $Z_E$ cancels the pinch-pole singularity and the cusp divergences (as well as the linear divergences if lattice regularization is employed) of the numerator in Eq.~(\ref{eq:quasi:def}), while leaving intact the local UV divergences arising from the gluon-Wilson line vertices at the endpoints of the bilinear operator. 
 If we work in the $\overline{\mathrm {MS}}$ scheme, the remaining local UV divergences in Eq.~(\ref{Eq:O1result}) can be removed by the following renormalization factor
\begin{align}
    \widetilde{Z}_1( \mu, \epsilon)&=1-\frac{\alpha_s}{4\pi} \left(
\beta_{0}-2C_A\right)\frac{1}{\epsilon_{\mathrm{UV}}},
\end{align}
so that the $\overline{\mathrm {MS}}$ renormalized quasi-TMDPDF becomes
\begin{align}\label{eq:MSbarunpol}
 &  x \t f_{1,\overline{\rm MS}}^{g} (x,\vec {b}_{\perp}^2,\mu, p^z) = \delta(1-x)+\frac{\alpha_{s}}{2 \pi}\left\{   -\left(\frac{1}{\epsilon_{\mathrm{IR}}}+\ln\frac{\mu^2 \vec{b}_{\perp}^2  e^{2\gamma_E}}{4} \right)x  P_{g  g}\right.\\ 
& \left.+\delta(1-x) C_{A} \bigg[\left(
\frac{\beta_{0}}{2C_A}-1\right)\ln\frac{\mu^2 \vec{b}_{\perp}^2  e^{2\gamma_E}}{4}-\frac{1}{2}\ln^2 ((p^z)^2 \vec{b}_{\perp}^2  e^{2\gamma_E}) +2\ln  ((p^z)^2 \vec{b}_{\perp}^2 e^{2\gamma_E})   -4\bigg] \right\}. \nn
\end{align}
%

However, when calculating the quasi-TMDPDFs on lattice, one has to choose a different renormalization scheme. In Refs.~\cite{Ji:2021uvr,Zhang:2022xuw}, two schemes have been proposed for the renormalization of the quark quasi-TMDPDF. They can be generalized to the gluon case by noting the different kinematic dependence and IR structure of the zero momentum matrix element. In the first scheme, we can do the renormalization by dividing by the matrix element in the rest frame in Eq.~(\ref{eq:quasi:def}) at small $\eta=\eta_0=(0,\vec{b}_{\perp,0},b^z_0)$,  
\beq\label{eq:restframeME1}
Z_{O,1}=\frac{1}{(P^0)^2}\tilde h_1(\eta=\eta_0,P^z=0).
\eeq
To simplify the calculation, we take $b_0^z=0$. However, the matrix element in Eq.~(\ref{eq:restframeME1}) contains IR divergences, which are identical to those in the local matrix element $\frac{1}{(P^+)^2}h(0)=\frac{1}{(P^+)^2}\langle P|F^{+i}(0)F^{+i}(0)|P\rangle$. Therefore, we can factorize the above renormalized quasi-TMDPDF to the standard TMDPDF normalized to $\frac{1}{(P^+)^2}h(0)$.
In the second scheme, 
we can divide by a ratio function formed by the straight line gauge link matrix elements
\beq
Z_{O,2}=\frac{1}{(P^0)^2}\frac{\tilde h_1^2(\eta_0/2,P^z=0)|_{\vec b_{\perp,0}=0}}{\tilde h_1(\eta_0,P^z=0)|_{\vec b_{\perp,0}=0}}.
\eeq

For illustration purposes, we will choose the first scheme in the discussion below.
The fully renormalized gluon quasi-TMDPDF is then given by
\beq
\tilde h_{1,R}(\eta,\eta_0,P^z)=Z_{O,1}^{-1}\tilde h_{1}(\eta,P^z),
\eeq
which can also be converted to the $\overline{\mathrm {MS}}$ scheme by multiplying a conversion factor:
\beq
\tilde h_{1,\overline{\rm MS}}(\eta, P^z,\mu)=\frac{1}{(P^0)^2}\tilde h_{1,R}^{\overline{\rm MS}}(\eta_0,P^z=0,\mu)\tilde h_{1,R}(\eta,\eta_0,P^z),
\eeq
where the result of the conversion factor up to one-loop is
\begin{align}
   &\frac{1}{(P^0)^2}\tilde h_{1,R}^{\overline{\rm MS}}(\eta_0,P^z=0,\mu)\nonumber\\
   =&1+\frac{\alpha_s }{2\pi}\bigg(\frac{\beta_0}{2}-C_A\bigg) \ln\frac{\mu^2 \vec{b}_{\perp,0}^2 e^{2\gamma_E}}{4} +\frac{\alpha_s}{\pi}\frac13 T_F n_f\bigg(\frac{1}{\epsilon_{\mathrm{IR}}}+\ln\frac{\mu_{\mathrm{IR}}^2 \vec{b}_{\perp,0}^2 e^{2\gamma_E}}{4}\bigg)+\mathcal{O}(\alpha_s^2).
\end{align}

From the above results, one can see that both the quasi-TMDPDF and the TMDPDF have identical collinear IR structures which enables a matching formula between them. The different UV structure can be understood as that
a large but finite momentum $P^z$ is employed in the quasi-TMDPDF while the physical TMDPDF requests $P^z$ go to infinity. 
In the case of TMDPDFs, the rapidity singularity appears in the physical TMDPDF because of
the infinitely long light-like Wilson line. One has to introduce a regulator, say, $\delta$, to regularize the
rapidity singularity. In contrast, no rapidity regulator is required in the quasi-TMDPDF because
the Wilson line is of infinite length and space-like. However, the quasi-TMDPDF depends on $\ln P^z$ so
that the $P^z\to \infty$ limit cannot be taken smoothly. Thus, a matching coefficient is needed to take the
$P^z\to \infty$ limit and compensate the difference at the large rapidity. Our observations are consistent
with previous discussions on quark quasi-TMDPDFs~\cite{Ebert:2019okf,Ebert:2022fmh} as well as on gluon quasi-TMDPDFs~\cite{Schindler:2022eva}.

One can derive a $P^z$ evolution equation for the quasi-TMDPDF
\begin{align}
P^z \frac{d}{d P^z} \ln \t f_{1,R}^{g} (x,\vec {b}_{\perp}^2,\mu, P^z)=K(\vec {b}_{\perp}^2,\mu)+{\cal F}\left(\frac{(P^z)^2}{\mu^{2}}\right),
\end{align}
where the subscript $R$ denotes the renormalized result, and the parton momentum has been expressed in terms of the hadron momentum as $p^z=x P^z$. Here the kernel $K(\vec {b}_{\perp}^2,\mu)$ is identical to the Collins-Soper kernel in the $\zeta$-evolution equation of the
gluon TMDPDFs. At one-loop, the Collins-Soper kernel takes the form
\begin{align}
K(\vec {b}_{\perp}^2,\mu)=-\frac{\alpha_{s}C_A}{ \pi}\ln\frac{\mu^2 \vec{b}_{\perp}^2  e^{2\gamma_E}}{4},
\end{align}
and $\cal F$ is independent of the transverse coordinate
\begin{align}
{\cal F}\left(\frac{(P^z)^2}{\mu^{2}}\right)=\frac{\alpha_{s}C_A}{ \pi}\left(-\ln \frac{(2xP^z)^2}{\mu^{2}}+2\right).
\end{align}
Note that the Collins-Soper kernel has been calculated to the four-loop order~\cite{Moult:2022xzt,Duhr:2022yyp}.

To explicitly express the difference between the gluon quasi-TMDPDF and TMDPDF, we form the following ratio between the $\overline{\mathrm {MS}}$ results
\begin{align}\label{eq:tmdratio}
 \frac{\t f_{1,\overline{\mathrm{MS}}}^{g} (x,\vec {b}_{\perp}^2, P^z)}{ f_{1,\overline{\mathrm{MS}}}^{g} (x,\vec {b}_{\perp}^2)}=&1+ \frac{\alpha_{s}C_A}{2 \pi}\left[-\ln\frac{\mu^2 \vec{b}_{\perp}^2  e^{2\gamma_E}}{4} \ln \frac{(2xP^z)^2}{\zeta}+\ln\frac{\mu^2 \vec{b}_{\perp}^2  e^{2\gamma_E}}{4} -\frac{1}{2}\ln^2 \frac{(2xP^z)^2}{\mu^{2}}\right.\nonumber\\&\left.+2\ln \frac{(2xP^z)^2}{\mu^{2}}+\frac{\pi^{2}}{12}-4\right].
\end{align}
As can be seen from the equation, apart from the $\ln\frac{\mu^2 \vec{b}_{\perp}^2  e^{2\gamma_E}}{4} \ln \frac{(2xP^z)^2}{\zeta}$ term that can be resummed by the $P^z$ evolution equation, there is an extra logarithmic term $C_A \ln\frac{\mu^2 \vec{b}_{\perp}^2  e^{2\gamma_E}}{4}$ which can be non-perturbative when $|\vec b_\perp|$ gets large (or $|\vec k_\perp|$ gets small), and thus potentially invalidates the existence of a factorization. In the case of quark quasi-TMDPDFs, this is cured by including a so-called reduced soft function, which serves the purpose of removing the rapidity scheme dependence. We can do the same here by introducing the reduced soft function for the gluon quasi-TMDPDFs, which can be extracted by studying the large rapidity behavior of the soft function in the off-light-cone regularization scheme~\cite{Ji:2020ect}
%
%
%
\begin{align}
S\left(\vec{b}_{\perp}^2, \mu, Y, Y^{\prime}\right)=\frac{\left\langle 0\left|\mathcal{W}_{+n_{Y^{\prime}}}\left(\vec{b}_{\perp}\right) \mathcal{W}_{+\bar{n}_Y}^{\dagger}\left(\vec{b}_{\perp}\right)\right| 0\right\rangle}{(N^2_{\mathrm{c}}-1) \sqrt{Z_{\mathrm{E}}^{\prime}} \sqrt{Z_{\mathrm{E}}}},
\end{align}
where the off-light-cone vectors $n_{Y^{\prime}}
=n-e^{-2Y^\prime} \bar{n}$ and $\bar{n}_{Y}
=\bar{n}-e^{-2Y} n$. $Z_{\mathrm{E}}$ and $Z_{\mathrm{E}}^{\prime}$ are introduced to remove the
pinch-pole singularities. Asymptotically, the off-light-cone soft function behaves like
\begin{align}
S\left(\vec{b}_{\perp}^2, \mu, Y, Y^{\prime}\right)=e^ {\left(Y+Y^{\prime}\right) K\left(\vec{b}_{\perp}^2, \mu\right)+\mathcal{D}\left(\vec{b}_{\perp}^2, \mu\right)}+\mathcal{O}\left(\mathrm{e}^{-\left(Y+Y^{\prime}\right)}\right),
\end{align}
from which we obtain the reduced soft function
\begin{align}
S_{r}\left(\vec{b}_{\perp}^2, \mu\right)=e^ {-\mathcal{D}\left(\vec{b}_{\perp}, \mu\right)},
\end{align}
with
\begin{align}
\mathcal{D}\left(\vec{b}_{\perp}^2, \mu\right)=\frac{\alpha_{s}C_A}{ \pi}\left(\frac{1}{\epsilon_{\mathrm{UV}}}+\ln\frac{\mu_{\mathrm{UV}}^2 \vec{b}_{\perp}^2  e^{2\gamma_E}}{4}\right)
\end{align}
at perturbative one-loop order. It exactly cancels the extra logarithmic term in Eq.~(\ref{eq:tmdratio}) if we set $\mu_{\rm UV}=\mu$.


With similar calculations, one can also derive the one-loop result of the polarized gluon quasi-TMDPDF $\tilde{g}_{1L}^g $, which can be obtained from the unpolarized one by replacing $P_{gg}$ with $\Delta P_{gg}$, e.g., in Eq.~\eqref{eq:MSbarunpol}.

It is worth mentioning at this stage that we have also calculated the mixing diagrams between gluons and quarks, which yield the same result for the quasi-TMDPDF and TMDPDF, in agreement with Ref.~\cite{Ebert:2022fmh}. Thus, such contributions do not affect the factorization to be discussed below.

\section{Quasi-TMDPDFs with $\delta$-regularization and scheme independence }
\label{sec:schemeindep}
Before presenting the factorization formula connecting the gluon quasi-TMDPDFs and TMDPDFs, we discuss the quasi-TMDPDF result with a different pinch-pole regulator and its implications. Besides the finite gauge link length regulator, we also perform a calculation with the $\delta$-regulator. In this regularization scheme, the length of the gauge link is infinite but the denominator of the gauge link propagator changes as $n\cdot k\pm i 0\to n\cdot k\pm i\delta$. There is no contribution from the transverse gauge link at infinity.
Because there is no gauge link interaction in Fig.~\ref{fig:gluonTMDNOGL}, the results are the same between different schemes. While for Fig.~\ref{fig:gluonTMDGL} ~(a) and (b), we get
  \begin{align}
x \tilde{f}_1^{g,(1)} &(x, \vec{b}_{\perp}^2, {p^z})|_{\ref{fig:gluonTMDGL}(a)+(b)}=  \frac{\alpha_s C_A}{2\pi }  \bigg\{ -\left(\frac{1}{\epsilon_{\mathrm{IR}}}+\ln\frac{\mu_{\mathrm{IR}}^2 \vec{b}_{\perp}^2 e^{2\gamma_E}}{4}\right)\bigg[\frac{x(1+x)}{1-x} \bigg]_+  \\
&+ \delta(1-x)      \bigg[ \frac{1}{2\epsilon_{\mathrm{UV}}}+\frac{1}{2}\ln\frac{\mu_{\mathrm{UV}}^2}{(2p^z)^2} +\frac52\ln ((p^z)^2 \vec{b}_{\perp}^2 e^{2\gamma_E})-\frac12\ln^2 ((p^z)^2 \vec{b}_{\perp}^2 e^{2\gamma_E})-4\bigg] \bigg\} . \nn
\end{align}
The result is independent of $\delta$ as expected. For Fig.~\ref{fig:gluonTMDGL}(c), one has
\begin{align}
  x \tilde{f}_1^{g,(1)} (x,\vec{b}_{\perp}^2,{p^z}, \delta)|_{\ref{fig:gluonTMDGL}(c)}=&\frac{\alpha_s}{2\pi}   C_A      \left(\frac{\pi}{\delta |\vec{b}_{\perp}|}+\frac{1}{\epsilon_{\mathrm{UV}}}   + \ln \frac{\mu_{\mathrm{UV}}^2 \vec{b}_{\perp}^2 e^{2\gamma_E}}{4} \right)  \delta(1-x), \label{eq:3c}
\end{align}
where the pinch-pole singularity is regularized by $1/\delta$. On the other hand, the one-loop correction to the factor $Z_E$ is
\begin{align}
    Z_E=1+\frac{\alpha_s}{\pi}C_A \frac{\pi}{\delta |\vec{b}_{\perp}|} .
\end{align}
Therefore, after $\sqrt{Z_E}$ subtraction, Eq.~\eqref{eq:3c} becomes
\begin{align}
   x \tilde{f}_1^{g,(1)} (x,\vec{b}_{\perp}^2,{p^z})|_{\ref{fig:gluonTMDGL}(c)}\to  &\frac{\alpha_s}{2\pi}   C_A      \left(\frac{1}{\epsilon_{\mathrm{UV}}}   + \ln \frac{\mu_{\mathrm{UV}}^2 \vec{b}_{\perp}^2 e^{2\gamma_E}}{4} \right)  \delta(1-x), \label{eq:3csub}
\end{align}
which means that the pinch-pole singularity is canceled, and there is no $\delta$-dependence in the subtracted result. 

For the unsubtracted soft function, we have
\begin{align}
&\frac{\alpha_s C_A}{\pi} \left(\frac{1}{\epsilon_{\mathrm{UV}}} +\ln \frac{\mu_{\mathrm{UV}}^2 \vec{b}_{\perp}^2 e^{2\gamma_E}}{4}  \right)(1-Y-Y')+\frac{\alpha_s}{2\pi}   C_A       \frac{\pi}{\delta}\sqrt{\frac{\bar n_Y'^2}{\vec{b}_{\perp}^2}}   +\frac{\alpha_s}{2\pi}   C_A      \frac{\pi}{\delta}\sqrt{\frac{n_Y^2}{\vec{b}_{\perp}^2}}. 
\end{align}
Subtracted with $\sqrt{Z_E Z_E'}$, one obtains 
\begin{align}
  &\frac{\alpha_s C_A}{\pi}\left( \frac{1}{\epsilon_{\mathrm{UV}}} +\ln\frac{\mu_{\mathrm{UV}}^2 \vec{b}_{\perp}^2 e^{2\gamma_E}}{4}\right) (1-Y-Y'),
\end{align}
in which the $\delta$-dependence cancels. So, the reduced soft function at one-loop level is
\begin{align}
 S_{r}^{(1)}\left({\vec{b}_{\perp}}^2, \mu\right)= -  \frac{\alpha_s C_A}{\pi}  \left(\frac{1}{\epsilon_{\mathrm{UV}}} +\ln\frac{\mu^2 \vec{b}_{\perp}^2 e^{2\gamma_E}}{4} \right),
\end{align}
which completely cancels the contribution in Eq.~\eqref{eq:3csub}.

From the above discussion, one can demonstrate that the one-loop results are independent of the pinch-pole regulator. This observation has the potential to greatly facilitate perturbative calculations, since we do not need to keep a finite gauge link length for perturbative calculations, which makes the latter very complicated. Instead, we can choose simpler regulators such as the $\delta$-regulator.

\section{Factorization of gluon quasi-TMDPDFs}
\label{sec:factor}

{ We are now ready to write down the matching formula between the gluon 
quasi-TMDPDFs and TMDPDFs as
\begin{align}
\t f_{1,R}^g\left(x, \vec{b}_{\perp}^2, \vec{b}_{\perp,0}^2, \zeta_{z}\right) S_{r}^{\frac{1}{2}}\left(\vec{b}_{\perp}^2, \mu\right)
&=H_R\left(\frac{\zeta_{z}}{\mu^{2}},\mu^2 \vec{b}_{\perp,0}^2\right) e^{\ln \frac{\zeta z}{\zeta} K\left(\vec{b}_{\perp}^2, \mu\right)} {\bar f}_1^g\left(x, \vec{b}_{\perp}^2, \mu, \zeta\right)+ p. c., \label{eq:fact} 
\end{align}
where $\t f_{1,R}^g$ denotes the renormalized quasi-TMDPDF matrix element using the method in Sec.~\ref{sec:gqtmdrenorm}. ${\bar f}_1^g=(P^+)^2 f_1^g/h(0)$ is the normalized $\overline{\rm MS}$ TMDPDF from which we can extract the physical TMDPDF $f_1^g$, $p. c.$ stands for power corrections. 
$S_r$ is the reduced soft function that can be calculated from a meson form factor~\cite{Ji:2020ect,Ji:2021znw,Deng:2022gzi} and converted to the $\overline{\rm MS}$ scheme. Here we assume it has been converted to the $\overline{\rm MS}$ scheme. $H_R$ is the hard matching kernel, which can be calculated perturbatively order by order. Up to the next-to-leading order, we have
\begin{align}
& H_{R}\left(\frac{\zeta_{z}}{\mu^{2}}, \vec{b}_{\perp,0}^2\right)= 1+
 \frac{\alpha_s C_A}{2\pi}\left( -\frac{1}{2}\ln^2 \frac{\zeta_z}{\mu^{2}}+2\ln \frac{\zeta_z}{\mu^{2}}-\frac{5}{6} \ln\frac{\mu^2 \vec{b}_{\perp,0}^2 e^{2\gamma_E}}{4}+\frac{\pi^{2}}{12}-4\right), 
\end{align}
where $\zeta_z\equiv (2 x P^z)^2$. 
The hard kernel satisfies the renormalization group equation
\begin{align}
&\mu \frac{d}{d \mu} \ln H_{R}\left(\frac{\zeta_{z}}{\mu^{2}}\right)=\Gamma_{\text {cusp }} \ln \frac{\zeta_{z}}{\mu^{2}}+\gamma_{C}+\gamma_R\, ,
\end{align}
where $\Gamma_{\mathrm{cusp}}^{(1)}=\frac{\alpha_s C_A}{\pi}$, $\gamma^{(1)}_{C}=-\frac{2\alpha_s C_A}{\pi}$ and $\gamma_R^{(1)}=-\frac56 \frac{\alpha_s C_A}{\pi}$. Note that the hard kernel may depend on the operator choice of the gluon quasi-TMDPDFs. Our result here is different from that in~\cite{Schindler:2022eva}. This is because, on the one hand, we have used different operators for the gluon quasi-TMDPDFs, and the corresponding gluon-Wilson-line vertices have different UV behavior, thus yielding different contributions to the matching kernel; on the other hand, our renormalization is carried out in the ratio scheme, which facilitates practical lattice calculations. 

For comparison, we also present the matching relation for the $\overline{\rm MS}$ renormalized quasi-TMDPDF, which is
\begin{align}
\t f_{1,\overline{\mathrm{MS}}}^g\left(x, \vec{b}_{\perp}^2, \mu, \zeta_{z}\right) S_{r}^{\frac{1}{2}}\left(\vec{b}_{\perp}^2, \mu\right)
&=H_{\overline{\mathrm{MS}}}\left(\frac{\zeta_{z}}{\mu^{2}}\right) e^{\ln \frac{\zeta z}{\zeta} K\left(\vec{b}_{\perp}^2, \mu\right)}  f_1^g\left(x, \vec{b}_{\perp}^2, \mu, \zeta\right)+ p. c., 
\end{align}
and the corresponding matching kernel reads
\begin{align}
& H_{\overline{\mathrm{MS}}}\left(\frac{\zeta_{z}}{\mu^{2}}\right)=1+
 \frac{\alpha_s C_A}{2\pi}\left( -\frac{1}{2}\ln^2 \frac{\zeta_z}{\mu^{2}}+2\ln \frac{\zeta_z}{\mu^{2}}+\frac{\pi^{2}}{12}-4\right), \label{eq:matchingMSbar}
\end{align}
and satisfies the renormalization group equation
\begin{align}
&\mu \frac{d}{d \mu} \ln H_{\overline{\mathrm{MS}}}\left(\frac{\zeta_{z}}{\mu^{2}}\right)=\Gamma_{\text {cusp }} \ln \frac{\zeta_{z}}{\mu^{2}}+\gamma_{C}.
\end{align}
}

We have also calculated the one-loop result of $g_{1L}^g$ and 
$\tilde{g}_{1L}^g$. We find that their results differ from $f_{1}^g$ and $\tilde{f}_{1}^{g}$ only by the splitting kernels. The splitting kernels are associated with the IR part. They are completely canceled in the matching. The factorization formulas for spin-dependent TMDs are in the same form as Eq.~\eqref{eq:fact}  (see, e.g., Ref.~\cite{Ebert:2022fmh}). Thus, the matching kernel for the polarized gluon quasi-TMDPDFs is the same as that for the unpolarized one. The matching coefficient for the linearly polarized gluon TMD at one-loop is also the same one~\cite{Zhao:2022qym}. In the case of quark quasi-TMDPDF, it has been observed that the matching coefficient for helicity and transversity distributions are equal to the one for the unpolarized TMDPDF~\cite{Ebert:2020gxr}. A similar factorization formula for the quark quasi-TMDPDFs has also been derived by employing the soft-collinear effective theory in Ref.~[27], and the matching coefficients are independent of the spin structure. We expect the same conclusion for the leading-twist gluon quasi-TMDPDFs. Of course, to confirm this, further investigations and calculations of the matching formula are required. We leave this to future work. 


\section{Conclusion}
\label{sec:concl}

In this paper, we have studied the lattice calculation strategy of gluon TMDPDFs. By taking the unpolarized and helicity gluon TMDPDFs as an example, we have performed a perturbative one-loop calculation for both the quasi-TMDPDFs and the TMDPDFs, and presented some lattice-friendly renormalization schemes for the former, which can be converted to the $\overline{\rm MS}$ scheme through a conversion factor. We then present a matching formula between the gluon quasi-TMDPDFs and TMDPDFs, which takes a similar form as in the quark case. Our results also indicate an independence of the regulator for the pinch-pole singularity. This observation has the potential to greatly facilitate calculations of the perturbative hard matching kernel. 

\acknowledgments

We thank Minhuan Chu, Jun Hua, Wei Wang, Fei Yao, Qi-An Zhang and Yong Zhao for useful discussions. RLZ is supported by NSFC under grant No.~12075124 and by Natural Science Foundation of Jiangsu under Grant No.~BK20211267. Y.J. is supported by the Collaborative Research
Center TRR110/2 funded by the Deutsche Forschungsgemeinschaft (DFG, German Research Foundation) under grant 409651613.  JHZ is supported in part
by National Natural Science Foundation of China under
grant No. 11975051, No. 1206113100. The work  of S.~Z. is supported by Jefferson Science Associates, LLC under  U.S. DOE Contract \# DE-AC05-06OR23177 and by U.S. DOE Grant \# DE-FG02-97ER41028.

\appendix
\section{Fourier transform integrals}
\label{Sec:FTint}
Some useful integrals for Fourier transform from transverse momentum to transverse coordinate are listed in the following:
\begin{align}
	 &\int d^{2-2\epsilon}k_{\perp}\frac{e^{-i\vec{k}_{\perp}\cdot \vec{b}_{\perp}}}{\vec{k}_{\perp}^2}= \pi \left(\frac{\vec{b}_{\perp}^2}{4\pi} \right)^{\epsilon}  \Gamma(-\epsilon)  \, , \nonumber\\
	 &\int d^{2}k_{\perp} \frac{\vec{k}_{\perp}^2}{\vec{k}_{\perp}^4+\Lambda^4}e^{-i\vec{k}_{\perp}\cdot \vec{b}_{\perp}} |_{\Lambda\to 0}= -\pi \ln\frac{\Lambda^2\vec{b}_{\perp}^2  e^{2\gamma_E}}{4}   \, , \nonumber\\
     &\int d^{2}k_{\perp} \frac{1}{\vec{k}_{\perp}^2+\Lambda^2} \ln \frac{\Lambda^2}{\vec{k}_{\perp}^2}e^{-i\vec{k}_{\perp}\cdot \vec{b}_{\perp}} |_{\Lambda\to 0}= -\pi \left( \frac12 \ln^2 \frac{ \Lambda^2\vec{b}_{\perp}^2  e^{2\gamma_E}}{4}+\frac{\pi^2}{3}\right)  \, , \nonumber\\	 
	 &\int d^{2-2\epsilon}k_{\perp}\frac{\ln \vec{k}_{\perp}^2}{\vec{k}_{\perp}^2} e^{-i\vec{k}_{\perp}\cdot \vec{b}_{\perp}}=-\pi \left(\frac{\vec{b}_{\perp}^2}{4\pi} \right)^{\epsilon}\Gamma(-\epsilon)\left[\gamma_E+\ln\frac{\vec{b}_{\perp}^2}{4}-\psi^{(0)}(-\epsilon)\right]\, ,
\end{align}
where $\psi^{(n)}$ is the polygamma function of order $n$.

\bibliographystyle{jhep}
\bibliography{ref}

\providecommand{\href}[2]{#2}\begingroup\raggedright\begin{thebibliography}{10}

\bibitem{Constantinou:2020hdm}
M.~Constantinou et~al., \emph{{Parton distributions and lattice-QCD
  calculations: Toward 3D structure}},
  \href{https://doi.org/10.1016/j.ppnp.2021.103908}{\emph{Prog. Part. Nucl.
  Phys.} {\bfseries 121} (2021) 103908}
  [\href{https://arxiv.org/abs/2006.08636}{{\ttfamily 2006.08636}}].

\bibitem{Collins:2011zzd}
J.~Collins, \emph{{Foundations of perturbative QCD}}, vol.~32, Cambridge
  University Press (11, 2013).

\bibitem{Echevarria:2011epo}
M.G.~Echevarria, A.~Idilbi and I.~Scimemi, \emph{{Factorization Theorem For
  Drell-Yan At Low $q_T$ And Transverse Momentum Distributions
  On-The-Light-Cone}},
  \href{https://doi.org/10.1007/JHEP07(2012)002}{\emph{JHEP} {\bfseries 07}
  (2012) 002} [\href{https://arxiv.org/abs/1111.4996}{{\ttfamily 1111.4996}}].

\bibitem{Echevarria:2012js}
M.G.~Echevarr\'\i{}a, A.~Idilbi and I.~Scimemi, \emph{{Soft and Collinear
  Factorization and Transverse Momentum Dependent Parton Distribution
  Functions}},
  \href{https://doi.org/10.1016/j.physletb.2013.09.003}{\emph{Phys. Lett. B}
  {\bfseries 726} (2013) 795}
  [\href{https://arxiv.org/abs/1211.1947}{{\ttfamily 1211.1947}}].

\bibitem{Bacchetta:2017gcc}
A.~Bacchetta, F.~Delcarro, C.~Pisano, M.~Radici and A.~Signori,
  \emph{{Extraction of partonic transverse momentum distributions from
  semi-inclusive deep-inelastic scattering, Drell-Yan and Z-boson production}},
  \href{https://doi.org/10.1007/JHEP06(2017)081}{\emph{JHEP} {\bfseries 06}
  (2017) 081} [\href{https://arxiv.org/abs/1703.10157}{{\ttfamily
  1703.10157}}].

\bibitem{Scimemi:2017etj}
I.~Scimemi and A.~Vladimirov, \emph{{Analysis of vector boson production within
  TMD factorization}},
  \href{https://doi.org/10.1140/epjc/s10052-018-5557-y}{\emph{Eur. Phys. J. C}
  {\bfseries 78} (2018) 89} [\href{https://arxiv.org/abs/1706.01473}{{\ttfamily
  1706.01473}}].

\bibitem{Bertone:2019nxa}
V.~Bertone, I.~Scimemi and A.~Vladimirov, \emph{{Extraction of unpolarized
  quark transverse momentum dependent parton distributions from
  Drell-Yan/Z-boson production}},
  \href{https://doi.org/10.1007/JHEP06(2019)028}{\emph{JHEP} {\bfseries 06}
  (2019) 028} [\href{https://arxiv.org/abs/1902.08474}{{\ttfamily
  1902.08474}}].

\bibitem{Scimemi:2019cmh}
I.~Scimemi and A.~Vladimirov, \emph{{Non-perturbative structure of
  semi-inclusive deep-inelastic and Drell-Yan scattering at small transverse
  momentum}}, \href{https://doi.org/10.1007/JHEP06(2020)137}{\emph{JHEP}
  {\bfseries 06} (2020) 137}
  [\href{https://arxiv.org/abs/1912.06532}{{\ttfamily 1912.06532}}].

\bibitem{Bacchetta:2019sam}
A.~Bacchetta, V.~Bertone, C.~Bissolotti, G.~Bozzi, F.~Delcarro, F.~Piacenza
  et~al., \emph{{Transverse-momentum-dependent parton distributions up to
  N$^{3}$LL from Drell-Yan data}},
  \href{https://doi.org/10.1007/JHEP07(2020)117}{\emph{JHEP} {\bfseries 07}
  (2020) 117} [\href{https://arxiv.org/abs/1912.07550}{{\ttfamily
  1912.07550}}].

\bibitem{Bacchetta:2020gko}
A.~Bacchetta, F.~Delcarro, C.~Pisano and M.~Radici, \emph{{The 3-dimensional
  distribution of quarks in momentum space}},
  \href{https://doi.org/10.1016/j.physletb.2022.136961}{\emph{Phys. Lett. B}
  {\bfseries 827} (2022) 136961}
  [\href{https://arxiv.org/abs/2004.14278}{{\ttfamily 2004.14278}}].

\bibitem{Zhu:2013yxa}
R.~Zhu, P.~Sun and F.~Yuan, \emph{{Low Transverse Momentum Heavy Quark Pair
  Production to Probe Gluon Tomography}},
  \href{https://doi.org/10.1016/j.physletb.2013.11.002}{\emph{Phys. Lett. B}
  {\bfseries 727} (2013) 474}
  [\href{https://arxiv.org/abs/1309.0780}{{\ttfamily 1309.0780}}].

\bibitem{Scarpa:2019fol}
F.~Scarpa, D.~Boer, M.G.~Echevarria, J.-P.~Lansberg, C.~Pisano and M.~Schlegel,
  \emph{{Studies of gluon TMDs and their evolution using quarkonium-pair
  production at the LHC}},
  \href{https://doi.org/10.1140/epjc/s10052-020-7619-1}{\emph{Eur. Phys. J. C}
  {\bfseries 80} (2020) 87} [\href{https://arxiv.org/abs/1909.05769}{{\ttfamily
  1909.05769}}].

\bibitem{Bacchetta:2022crh}
A.~Bacchetta, F.G.~Celiberto and M.~Radici, \emph{{Unveiling the proton
  structure via transverse-momentum-dependent gluon distributions}},
  \href{https://doi.org/10.31349/SuplRevMexFis.3.0308108}{\emph{Rev. Mex. Fis.
  Suppl.} {\bfseries 3} (2022) 0308108}
  [\href{https://arxiv.org/abs/2206.07815}{{\ttfamily 2206.07815}}].

\bibitem{Hagler:2009mb}
P.~Hagler, B.U.~Musch, J.W.~Negele and A.~Schafer, \emph{{Intrinsic quark
  transverse momentum in the nucleon from lattice QCD}},
  \href{https://doi.org/10.1209/0295-5075/88/61001}{\emph{EPL} {\bfseries 88}
  (2009) 61001} [\href{https://arxiv.org/abs/0908.1283}{{\ttfamily
  0908.1283}}].

\bibitem{Musch:2011er}
B.U.~Musch, P.~Hagler, M.~Engelhardt, J.W.~Negele and A.~Schafer, \emph{{Sivers
  and Boer-Mulders observables from lattice QCD}},
  \href{https://doi.org/10.1103/PhysRevD.85.094510}{\emph{Phys. Rev. D}
  {\bfseries 85} (2012) 094510}
  [\href{https://arxiv.org/abs/1111.4249}{{\ttfamily 1111.4249}}].

\bibitem{Engelhardt:2015xja}
M.~Engelhardt, P.~H\"agler, B.~Musch, J.~Negele and A.~Sch\"afer,
  \emph{{Lattice QCD study of the Boer-Mulders effect in a pion}},
  \href{https://doi.org/10.1103/PhysRevD.93.054501}{\emph{Phys. Rev. D}
  {\bfseries 93} (2016) 054501}
  [\href{https://arxiv.org/abs/1506.07826}{{\ttfamily 1506.07826}}].

\bibitem{Ji:2014hxa}
X.~Ji, P.~Sun, X.~Xiong and F.~Yuan, \emph{{Soft factor subtraction and
  transverse momentum dependent parton distributions on the lattice}},
  \href{https://doi.org/10.1103/PhysRevD.91.074009}{\emph{Phys. Rev. D}
  {\bfseries 91} (2015) 074009}
  [\href{https://arxiv.org/abs/1405.7640}{{\ttfamily 1405.7640}}].

\bibitem{Ji:2018hvs}
X.~Ji, L.-C.~Jin, F.~Yuan, J.-H.~Zhang and Y.~Zhao, \emph{{Transverse momentum
  dependent parton quasidistributions}},
  \href{https://doi.org/10.1103/PhysRevD.99.114006}{\emph{Phys. Rev. D}
  {\bfseries 99} (2019) 114006}
  [\href{https://arxiv.org/abs/1801.05930}{{\ttfamily 1801.05930}}].

\bibitem{Ji:2019sxk}
X.~Ji, Y.~Liu and Y.-S.~Liu, \emph{{TMD soft function from large-momentum
  effective theory}},
  \href{https://doi.org/10.1016/j.nuclphysb.2020.115054}{\emph{Nucl. Phys. B}
  {\bfseries 955} (2020) 115054}
  [\href{https://arxiv.org/abs/1910.11415}{{\ttfamily 1910.11415}}].

\bibitem{Ji:2019ewn}
X.~Ji, Y.~Liu and Y.-S.~Liu, \emph{{Transverse-momentum-dependent parton
  distribution functions from large-momentum effective theory}},
  \href{https://doi.org/10.1016/j.physletb.2020.135946}{\emph{Phys. Lett. B}
  {\bfseries 811} (2020) 135946}
  [\href{https://arxiv.org/abs/1911.03840}{{\ttfamily 1911.03840}}].

\bibitem{Ji:2020jeb}
X.~Ji, Y.~Liu, A.~Sch\"afer and F.~Yuan, \emph{{Single Transverse-Spin
  Asymmetry and Sivers Function in Large Momentum Effective Theory}},
  \href{https://doi.org/10.1103/PhysRevD.103.074005}{\emph{Phys. Rev. D}
  {\bfseries 103} (2021) 074005}
  [\href{https://arxiv.org/abs/2011.13397}{{\ttfamily 2011.13397}}].

\bibitem{Ebert:2018gzl}
M.A.~Ebert, I.W.~Stewart and Y.~Zhao, \emph{{Determining the Nonperturbative
  Collins-Soper Kernel From Lattice QCD}},
  \href{https://doi.org/10.1103/PhysRevD.99.034505}{\emph{Phys. Rev. D}
  {\bfseries 99} (2019) 034505}
  [\href{https://arxiv.org/abs/1811.00026}{{\ttfamily 1811.00026}}].

\bibitem{Ebert:2019okf}
M.A.~Ebert, I.W.~Stewart and Y.~Zhao, \emph{{Towards Quasi-Transverse Momentum
  Dependent PDFs Computable on the Lattice}},
  \href{https://doi.org/10.1007/JHEP09(2019)037}{\emph{JHEP} {\bfseries 09}
  (2019) 037} [\href{https://arxiv.org/abs/1901.03685}{{\ttfamily
  1901.03685}}].

\bibitem{Ebert:2019tvc}
M.A.~Ebert, I.W.~Stewart and Y.~Zhao, \emph{{Renormalization and Matching for
  the Collins-Soper Kernel from Lattice QCD}},
  \href{https://doi.org/10.1007/JHEP03(2020)099}{\emph{JHEP} {\bfseries 03}
  (2020) 099} [\href{https://arxiv.org/abs/1910.08569}{{\ttfamily
  1910.08569}}].

\bibitem{Shanahan:2019zcq}
P.~Shanahan, M.L.~Wagman and Y.~Zhao, \emph{{Nonperturbative renormalization of
  staple-shaped Wilson line operators in lattice QCD}},
  \href{https://doi.org/10.1103/PhysRevD.101.074505}{\emph{Phys. Rev. D}
  {\bfseries 101} (2020) 074505}
  [\href{https://arxiv.org/abs/1911.00800}{{\ttfamily 1911.00800}}].

\bibitem{Ebert:2020gxr}
M.A.~Ebert, S.T.~Schindler, I.W.~Stewart and Y.~Zhao, \emph{{One-loop Matching
  for Spin-Dependent Quasi-TMDs}},
  \href{https://doi.org/10.1007/JHEP09(2020)099}{\emph{JHEP} {\bfseries 09}
  (2020) 099} [\href{https://arxiv.org/abs/2004.14831}{{\ttfamily
  2004.14831}}].

\bibitem{Vladimirov:2020ofp}
A.A.~Vladimirov and A.~Sch\"afer, \emph{{Transverse momentum dependent
  factorization for lattice observables}},
  \href{https://doi.org/10.1103/PhysRevD.101.074517}{\emph{Phys. Rev. D}
  {\bfseries 101} (2020) 074517}
  [\href{https://arxiv.org/abs/2002.07527}{{\ttfamily 2002.07527}}].

\bibitem{Ji:2021uvr}
Y.~Ji, J.-H.~Zhang, S.~Zhao and R.~Zhu, \emph{{Renormalization and mixing of
  staple-shaped Wilson line operators on the lattice revisited}},
  \href{https://doi.org/10.1103/PhysRevD.104.094510}{\emph{Phys. Rev. D}
  {\bfseries 104} (2021) 094510}
  [\href{https://arxiv.org/abs/2104.13345}{{\ttfamily 2104.13345}}].

\bibitem{Zhang:2022xuw}
{\scshape [Lattice Parton Collaboration (LPC)]} collaboration,
  \emph{{Renormalization of Transverse-Momentum-Dependent Parton Distribution
  on the Lattice}},
  \href{https://doi.org/10.1103/PhysRevLett.129.082002}{\emph{Phys. Rev. Lett.}
  {\bfseries 129} (2022) 082002}
  [\href{https://arxiv.org/abs/2205.13402}{{\ttfamily 2205.13402}}].

\bibitem{Ebert:2022fmh}
M.A.~Ebert, S.T.~Schindler, I.W.~Stewart and Y.~Zhao, \emph{{Factorization
  connecting continuum \& lattice TMDs}},
  \href{https://doi.org/10.1007/JHEP04(2022)178}{\emph{JHEP} {\bfseries 04}
  (2022) 178} [\href{https://arxiv.org/abs/2201.08401}{{\ttfamily
  2201.08401}}].

\bibitem{Ji:2013dva}
X.~Ji, \emph{{Parton Physics on a Euclidean Lattice}},
  \href{https://doi.org/10.1103/PhysRevLett.110.262002}{\emph{Phys. Rev. Lett.}
  {\bfseries 110} (2013) 262002}
  [\href{https://arxiv.org/abs/1305.1539}{{\ttfamily 1305.1539}}].

\bibitem{Ji:2014gla}
X.~Ji, \emph{{Parton Physics from Large-Momentum Effective Field Theory}},
  \href{https://doi.org/10.1007/s11433-014-5492-3}{\emph{Sci. China Phys. Mech.
  Astron.} {\bfseries 57} (2014) 1407}
  [\href{https://arxiv.org/abs/1404.6680}{{\ttfamily 1404.6680}}].

\bibitem{Ji:2020ect}
X.~Ji, Y.-S.~Liu, Y.~Liu, J.-H.~Zhang and Y.~Zhao, \emph{{Large-momentum
  effective theory}},
  \href{https://doi.org/10.1103/RevModPhys.93.035005}{\emph{Rev. Mod. Phys.}
  {\bfseries 93} (2021) 035005}
  [\href{https://arxiv.org/abs/2004.03543}{{\ttfamily 2004.03543}}].

\bibitem{LatticeParton:2020uhz}
{\scshape Lattice Parton} collaboration, \emph{{Lattice-QCD Calculations of TMD
  Soft Function Through Large-Momentum Effective Theory}},
  \href{https://doi.org/10.22323/1.396.0477}{\emph{Phys. Rev. Lett.} {\bfseries
  125} (2020) 192001} [\href{https://arxiv.org/abs/2005.14572}{{\ttfamily
  2005.14572}}].

\bibitem{Li:2021wvl}
Y.~Li et~al., \emph{{Lattice QCD Study of Transverse-Momentum Dependent Soft
  Function}}, \href{https://doi.org/10.1103/PhysRevLett.128.062002}{\emph{Phys.
  Rev. Lett.} {\bfseries 128} (2022) 062002}
  [\href{https://arxiv.org/abs/2106.13027}{{\ttfamily 2106.13027}}].

\bibitem{LPC:2022ibr}
{\scshape LPC} collaboration, \emph{{Nonperturbative determination of the
  Collins-Soper kernel from quasitransverse-momentum-dependent wave
  functions}}, \href{https://doi.org/10.1103/PhysRevD.106.034509}{\emph{Phys.
  Rev. D} {\bfseries 106} (2022) 034509}
  [\href{https://arxiv.org/abs/2204.00200}{{\ttfamily 2204.00200}}].

\bibitem{Shanahan:2020zxr}
P.~Shanahan, M.~Wagman and Y.~Zhao, \emph{{Collins-Soper kernel for TMD
  evolution from lattice QCD}},
  \href{https://doi.org/10.1103/PhysRevD.102.014511}{\emph{Phys. Rev. D}
  {\bfseries 102} (2020) 014511}
  [\href{https://arxiv.org/abs/2003.06063}{{\ttfamily 2003.06063}}].

\bibitem{Shanahan:2021tst}
P.~Shanahan, M.~Wagman and Y.~Zhao, \emph{{Lattice QCD calculation of the
  Collins-Soper kernel from quasi-TMDPDFs}},
  \href{https://doi.org/10.1103/PhysRevD.104.114502}{\emph{Phys. Rev. D}
  {\bfseries 104} (2021) 114502}
  [\href{https://arxiv.org/abs/2107.11930}{{\ttfamily 2107.11930}}].

\bibitem{Schlemmer:2021aij}
M.~Schlemmer, A.~Vladimirov, C.~Zimmermann, M.~Engelhardt and A.~Sch\"afer,
  \emph{{Determination of the Collins-Soper Kernel from Lattice QCD}},
  \href{https://doi.org/10.1007/JHEP08(2021)004}{\emph{JHEP} {\bfseries 08}
  (2021) 004} [\href{https://arxiv.org/abs/2103.16991}{{\ttfamily
  2103.16991}}].

\bibitem{Schindler:2022eva}
S.T.~Schindler, I.W.~Stewart and Y.~Zhao, \emph{{One-loop matching for gluon
  lattice TMDs}}, \href{https://doi.org/10.1007/JHEP08(2022)084}{\emph{JHEP}
  {\bfseries 08} (2022) 084}
  [\href{https://arxiv.org/abs/2205.12369}{{\ttfamily 2205.12369}}].

\bibitem{Mulders:2000sh}
P.J.~Mulders and J.~Rodrigues, \emph{{Transverse momentum dependence in gluon
  distribution and fragmentation functions}},
  \href{https://doi.org/10.1103/PhysRevD.63.094021}{\emph{Phys. Rev. D}
  {\bfseries 63} (2001) 094021}
  [\href{https://arxiv.org/abs/hep-ph/0009343}{{\ttfamily hep-ph/0009343}}].

\bibitem{Meissner:2007rx}
S.~Meissner, A.~Metz and K.~Goeke, \emph{{Relations between generalized and
  transverse momentum dependent parton distributions}},
  \href{https://doi.org/10.1103/PhysRevD.76.034002}{\emph{Phys. Rev. D}
  {\bfseries 76} (2007) 034002}
  [\href{https://arxiv.org/abs/hep-ph/0703176}{{\ttfamily hep-ph/0703176}}].

\bibitem{Belitsky:2002sm}
A.V.~Belitsky, X.~Ji and F.~Yuan, \emph{{Final state interactions and gauge
  invariant parton distributions}},
  \href{https://doi.org/10.1016/S0550-3213(03)00121-4}{\emph{Nucl. Phys. B}
  {\bfseries 656} (2003) 165}
  [\href{https://arxiv.org/abs/hep-ph/0208038}{{\ttfamily hep-ph/0208038}}].

\bibitem{Echevarria:2015usa}
M.G.~Echevarria, I.~Scimemi and A.~Vladimirov, \emph{{Transverse momentum
  dependent fragmentation function at
  next-to\textendash{}next-to\textendash{}leading order}},
  \href{https://doi.org/10.1103/PhysRevD.93.011502}{\emph{Phys. Rev. D}
  {\bfseries 93} (2016) 011502}
  [\href{https://arxiv.org/abs/1509.06392}{{\ttfamily 1509.06392}}].

\bibitem{Echevarria:2015byo}
M.G.~Echevarria, I.~Scimemi and A.~Vladimirov, \emph{{Universal transverse
  momentum dependent soft function at NNLO}},
  \href{https://doi.org/10.1103/PhysRevD.93.054004}{\emph{Phys. Rev. D}
  {\bfseries 93} (2016) 054004}
  [\href{https://arxiv.org/abs/1511.05590}{{\ttfamily 1511.05590}}].

\bibitem{Zhang:2018diq}
J.-H.~Zhang, X.~Ji, A.~Sch\"afer, W.~Wang and S.~Zhao, \emph{{Accessing Gluon
  Parton Distributions in Large Momentum Effective Theory}},
  \href{https://doi.org/10.1103/PhysRevLett.122.142001}{\emph{Phys. Rev. Lett.}
  {\bfseries 122} (2019) 142001}
  [\href{https://arxiv.org/abs/1808.10824}{{\ttfamily 1808.10824}}].

\bibitem{Wang:2019tgg}
W.~Wang, J.-H.~Zhang, S.~Zhao and R.~Zhu, \emph{{Complete matching for
  quasidistribution functions in large momentum effective theory}},
  \href{https://doi.org/10.1103/PhysRevD.100.074509}{\emph{Phys. Rev. D}
  {\bfseries 100} (2019) 074509}
  [\href{https://arxiv.org/abs/1904.00978}{{\ttfamily 1904.00978}}].

\bibitem{Moult:2022xzt}
I.~Moult, H.X.~Zhu and Y.J.~Zhu, \emph{{The four loop QCD rapidity anomalous
  dimension}}, \href{https://doi.org/10.1007/JHEP08(2022)280}{\emph{JHEP}
  {\bfseries 08} (2022) 280}
  [\href{https://arxiv.org/abs/2205.02249}{{\ttfamily 2205.02249}}].

\bibitem{Duhr:2022yyp}
C.~Duhr, B.~Mistlberger and G.~Vita, \emph{{Four-Loop Rapidity Anomalous
  Dimension and Event Shapes to Fourth Logarithmic Order}},
  \href{https://doi.org/10.1103/PhysRevLett.129.162001}{\emph{Phys. Rev. Lett.}
  {\bfseries 129} (2022) 162001}
  [\href{https://arxiv.org/abs/2205.02242}{{\ttfamily 2205.02242}}].

\bibitem{Ji:2021znw}
X.~Ji and Y.~Liu, \emph{{Computing light-front wave functions without
  light-front quantization: A large-momentum effective theory approach}},
  \href{https://doi.org/10.1103/PhysRevD.105.076014}{\emph{Phys. Rev. D}
  {\bfseries 105} (2022) 076014}
  [\href{https://arxiv.org/abs/2106.05310}{{\ttfamily 2106.05310}}].

\bibitem{Deng:2022gzi}
Z.-F.~Deng, W.~Wang and J.~Zeng, \emph{{Transverse-momentum-dependent wave
  functions and soft functions at one-loop in large momentum effective
  theory}}, \href{https://doi.org/10.1007/JHEP09(2022)046}{\emph{JHEP}
  {\bfseries 09} (2022) 046}
  [\href{https://arxiv.org/abs/2207.07280}{{\ttfamily 2207.07280}}].

\bibitem{Zhao:2022qym}
S.~Zhao, \emph{{Counting linearly polarized gluons with lattice QCD}},
  \href{https://arxiv.org/abs/2212.00825}{{\ttfamily 2212.00825}}.

\end{thebibliography}\endgroup

\end{document}